\documentclass[%
 aip,
 amsmath,amssymb,
 reprint,%
]{revtex4-2}

\usepackage{graphicx}
\usepackage{dcolumn}
\usepackage{bm}
\usepackage{physics}
\usepackage{float}
\usepackage[T1]{fontenc}
\usepackage{amsmath}
\usepackage{subcaption}
\usepackage{color}
\newcommand{\parl}{\parallel}
\usepackage{verbatim}

\usepackage[utf8]{inputenc}
\usepackage[T1]{fontenc}
\usepackage{mathptmx}
\usepackage{etoolbox}
\def\UOB{Institute of Mathematics and Physics, Technical University of Bydgoszcz, Al. Prof. S. Kaliskiego 7, 85-789 Bydgoszcz, Poland}
\def\UCI{Department of Chemistry, University of California, Irvine, Irvine, California, 92697, USA}
\makeatletter
\def\@email#1#2{%
 \endgroup
 \patchcmd{\titleblock@produce}
  {\frontmatter@RRAPformat}
  {\frontmatter@RRAPformat{\produce@RRAP{*#1\href{mailto:#2}{#2}}}\frontmatter@RRAPformat}
  {}{}
}%
\makeatother
\begin{document}

\preprint{AIP/123-QED}

\title{Two-Photon Absorption in Silicon Using the Real Density Matrix Approach}
\author{David Ziemkiewicz}
\affiliation{\UOB}

\author{David Knez}
\author{Evan P. Garcia}
\affiliation{\UCI}
\author{Sylwia Zieli\'{n}ska-Raczy\'{n}ska}%
\author{Gerard Czajkowski}
\affiliation{\UOB}

\author{Alessandro Salandrino}
\affiliation{%
 Department of Electrical Engineering, University of Kansas, Lawrence, Kansas, 66045
}%

\author{Sergey S. Kharintsev}
\author{Aleksei I. Noskov}
\affiliation{Institute of Physics, Kazan Federal University, Kazan 420008, Russia}

\author{Eric O. Potma}
\affiliation{\UCI}
\author{Dmitry A. Fishman}%
 \email{dmitryf@uci.edu}
\affiliation{\UCI}%

\date{\today}

\begin{abstract}
Two-photon absorption in indirect gap semiconductors is an frequently encountered, but not well-understood phenomenon. To address this, the real-density matrix approach (RDMA) is applied to describe two-photon absorption in silicon through the excitonic response to the interacting fields. This approach produces an analytical expression for the dispersion of the two-photon absorption coefficient for indirect-gap materials, and can be used to explain trends in reported experimental data for bulk silicon both old and new with minimal fitting.
\end{abstract}

\maketitle

\section{\label{sec:intro}Introduction}
Two-photon absorption (2PA) is a nonlinear optical phenomenon in which the combined energy carried by two incident photons is near-instantaneously absorbed by a material. Because of its nonlinear intensity dependence, the 2PA process is spatially confined when focused light is used, thus making it possible to use the effect as a local probe in the material for imaging~\cite{Denk1990}, microfabrication~\cite{Maruo1997,kawata2001} or three-dimensional data storage~\cite{parthenopoulos1989}. In semiconducting materials, the 2PA phenomenon produces charge carriers than can be read out electronically, a mechanism that has been used extensively for measuring the envelopes of ultrashort optical pulses.~\cite{Rudolph1997,Lochbrunner2000} The use of semiconductors as 2PA detectors is particularly attractive in the case of non-degenerate two-photon absorption (NTA), in which case the energies of $\hbar\omega_a$ and $\hbar\omega_b$ of the two photons in the interaction can be vastly different, enabling the registration of individual photon energies that are well below the bandgap energy of the semiconducting material. For instance, the NTA process has made it possible to detect mid-infrared (MIR) photons with wide-bandgap semiconductors such as GaAs~\cite{fishman_sensitive_2011} and Si~\cite{fang_highly_2020}, thereby circumventing thermal noise issues that have plagued traditional MIR photodetectors based on low-energy bandgap materials. Moreover, the NTA principle has also been shown to enable rapid MIR imaging with high-definition visible/near-IR cameras~\cite{knez2020,potma_high-speed_2021,liu_spatial_2022}, thereby overcoming several persistent limitations of conventional MIR cameras.

The recent developments in NTA-based imaging underline the potential of this nascent detection technology, particularly as it concerns the use of Si-based cameras for versatile imaging in the MIR range.~\cite{knez2020,liu_spatial_2022} From a commercial point of view, silicon is an attractive material because of its ubiquity as a feedstock and the mature technology for Si-device fabrication. On the other hand, silicon is an indirect semiconductor that has a much lower 2PA absorption coefficient compared to direct semiconductor in the same spectral range. The main reason for silicon's lower performance as a 2PA material is the additional involvement of lattice phonons for providing the momentum needed in the indirect transition, which lowers the transition probability. 

Improving the 2PA response of Si-based photodectectors requires a better understanding of the indirect two-photon transition in silicon. Reasonable progress has been made in describing degenerate and non-degenerate two-photon absorption in \emph{direct} bandgap semiconductors. Current models compute the 2PA transition rate in such materials through a second-order perturbation of stationary material states~\cite{wherrett_scaling_1984,sheik-bahae_dispersion_1991,hutchings_nondegenerate_1992}, or via evolving Volkov-type wavefunctions in the context of first-order perturbation theory.~\cite{jalbert_dressed_1986}  These approaches have successfully produced expressions that generally predict the energy scaling of the two-photon absorption coefficient in direct bandgap materials, including the expected enhancement of the 2PA effect in the case of extremely non-degenerate photon energies.~\cite{fishman_sensitive_2011} Existing models are flexible with respect to the input band structure, but do not incorporate dephasing mechanisms explicitly, and extra care must be taken to account for material anisotropy.~\cite{hutchings_theory_1994} Alternative formulations include the use of semiconductor Bloch equations, which have been evaluated in the framework of the density operator that is better equipped to account for relaxation and dephasing mechanisms.~\cite{aversa1994,aversa_nonlinear_1995,hannes_higher-order_2019}

The description of 2PA in \emph{indirect} semiconductors such as silicon has so far relied heavily on the theoretical models developed for direct bandgap materials. For instance, to account for the interaction vertex with lattice phonons, higher order perturbative expansions of stationary~\cite{dinu_dispersion_2003,dinu_third-order_2003} or dressed states~\cite{hayat_infrared_2008} have been developed to derive expressions for the nonlinear absorption coefficient. Although such models reproduce the general behaviour of measured two-photon absorption coefficients, quantitative matching between experiment and theory has proven more challenging. Recently, Faryadraz et al. obtained a semi-empirical scaling law for NTA coefficients, which was compared with experimental data.~\cite{faryadras_non-degenerate_2021}. While semi-empirical models can be useful for gaining mechanistic insights, a more complete theory of two-photon absorption in indirect bandgap materials is needed to quantitatively predict the scaling of 2PA coefficients over a wide range of the energy ratio $\hbar\omega_b/\hbar\omega_a$. 

In this work, we advance the theoretical description of the two-photon absorption process in silicon through the application of the real density matrix approach (RDMA). This approach has been successful in describing linear and nonlinear optical properties of semiconductors in terms of Rydberg excitons for the case of one-photon excitation \cite{zeli_nonlinear_2019,morin_selfkerr_2022}. The RDMA method allows the use of a small number of well-known parameters (e.g., effective masses, gap energy, dielectric constant) to derive expressions for the optical response of the material. Here, we adapt the RDMA for the case of two-photon excitation in semiconductors, extending it for the case of silicon to include both excitonic and continuum states, as well as lattice phonons for momentum matching. Using this approach, we derive analytical expressions for the two-photon coefficient while including dephasing times and anisotropy of the material under study. We compare the energy scaling of the predicted NTA coefficient with published experimental data, supplemented with new experimental data, and demonstrate excellent quantitative agreement over a broad range of $\hbar\omega_b/\hbar\omega_a$ dispersion values.

\section{Theory}

\subsection{Real density matrix approach}\label{sec_RDMA}
In this section, we briefly review the basic principles of the RDMA method in the context of linear and nonlinear optical excitations in semiconductors. In general, the optical response of semiconductors to incoming electromagnetic (EM) waves can be described in terms of the correlations in a many-body system (semiconductor) caused by the interactions between the EM fields, electrons, and holes (quasiparticles). A variety of theoretical methods has been used to describe the response, including the Green function approach~\cite{Haug,Bechstedt} and the RDMA. The latter, also known as coherent wave theory or the band edge equations, was developed in works by Stahl, Balslev, Axt, Zimmermann and others, see for instance Refs.~\cite{Stahl84,RivistaGC}. 

In this work, we have chosen the RDMA to take advantage of its formulation in the real
space. The method provides a direct relation between the density matrices and relevant observables, allowing an easy comparison between experimental and theoretical results. In general, the Coulomb interaction between the carriers in a many-body system produces an infinite hierarchy of evolution equations for $n-$point density matrices. The lowest level consists of two-point density matrices, which describe the interband transitions, between the valence and the conduction band, as well as the
intraband transitions. Two-point density matrices are directly related to measurable quantities such as polarization and carrier densities, whereas higher order correlations are related, for example, to the formation of biexcitons.~\cite{AxtStahl94b}

Below we restrict our model to two-point density matrices. The basic equations of the chosen approach are called the constitutive equations, or band-edge equations, and will be
applied to describe the 2PA processes. We follow the description developed in Ref.~\cite{RivistaGC}.
The RDMA takes into account the following contributions:
a) the electron-hole interaction;
b) the dipole interaction between the electron-hole pairs and the
electromagnetic field;
c) the particle-surface interaction;
and d) effects of external fields. We consider a semiconductor in the real space representation,
characterized by a number of valence- and conduction bands. Electrons at site $j$ in the conduction band are described by fermion operators $\hat{c}^{c\dag}_{j} (\hat{c}^c_{{j}})$, which
correspond to  the creation (annihilation) operators. Similarly, operators $\hat{d}^{v\dag}_{{j}} (\hat{d}^v_{{j}})$ are creation (annihilation) operators for holes in valence bands at site $j$.
In the case of direct interband transitions, the Hamilton operator in our model consists of three parts
\begin{equation}\label{2quanhamiltonian}
H=H_0+H_{em}+{H}_C.
\end{equation}
\noindent The term $ {H}_0$ idnicates the one-particle Bloch states in the conduction and valence bands, and describes the intraband transport processes:
\begin{equation}
 { H}_0=\sum_{j\ell}\left(\sum\limits_c T^c_{\ell j}\hat{c}_{\ell}^{c\dag}\hat{c}_j^c
-\sum\limits_v T^v_{j\ell
}\hat{d}_{\ell}^{v\dag}\hat{d}_j^v\right),
\end{equation}
\noindent where the transfer matrices $T_{ij}^{c,v}$ are given in
terms of the band eigenvalues $E_{c,v}(k)$ as:
\begin{equation}
T_{ij}^c=\frac{1}{N}\sum\limits_k e^{ik(R_i-R_j)}E_c(k),
\end{equation}
\noindent with a similar expression for the valence band transfer
matrix $T_{ij}^v$. The operator $H_{em}$ describes the interaction
with the electromagnetic field:
\begin{equation}
H_{em}=-\sum\limits_{j\ell}{\bf E}_{j\ell}\cdot\sum_{cv}{\bf
M}^{cv*}_{j\ell}\hat{d}^v_{\ell}\hat{c}^c_j+{\rm h.c.},
\end{equation}
\noindent and ${\bf M}^{cv}_{j\ell}$ is the interband-dipole
matrix element between Wannier states, which can be cast in the following
form
\begin{equation}\label{dipolematrixelement}
{\bf M}^{cv}_{j\ell}=\frac{1}{N}\sum_k e^{ik(R_j-R_{\ell})}{\bf
M}_{cv}({\bf k}),
\end{equation}
\noindent in terms of the dipole matrix element between Bloch
functions
\begin{equation}
{\bf M}_{cv}({\bf k})=-e\langle \psi_c({\bf k})\vert{\bf
r}\vert\psi_v({\bf k})\rangle.
\end{equation}
\noindent
The electric field ${\bf E}_{j\ell}$ is considered at a mean point between the sites $j$ and $\ell$ of the lattice; below this position is assumed to coincide with the exciton center-of-mass. In the following, we assume the band structure is known from specific calculations.

The carrier interactions are described by the term ${H}_C$ of the Hamiltonian:
\begin{equation}\label{HC}
{ H}_C=\frac{1}{2}\sum\limits_{i\neq
j}V_{ij}(\hat{n}_i-\hat{h}_i)(\hat{n}_j-\hat{h}_j),
\end{equation}
\noindent where the electron-hole interaction is screened by a
background dielectric constant $\epsilon_b$
\begin{equation}\label{Coulombij}
V_{ij}=\frac{e^2}{4\pi\epsilon_0\epsilon_b\left|\textbf{r}_i-\textbf{r}_j\right|},
\end{equation}
\noindent
We also define the electron and hole occupation numbers as:
\begin{equation}
\hat{n}_j=\sum\limits_c \hat{c}_j^{c\dag}\hat{c}_j^c,\quad
\hat{h}_j=\sum\limits_v \hat{d}^{v\dag}_j\,\hat{d}^v_j.
\end{equation}

The physical quantities relevant for finding the optical properties can be expressed in terms of mean values of the following pair operators
\begin{eqnarray}
&&\hbox{excitonic transition density amplitude:}\nonumber\\
&&Y_{12}^{\alpha\,b}=\langle\hat{Y}_{12}^{\alpha\,b}\rangle
=\langle\hat{d}^{\alpha}_1\,\hat{c}^b_2\rangle,\nonumber\\
&&\hbox{electron density:}\quad
C_{12}^{ab}=\langle\hat{C}^{ab}_{12}\rangle=\langle{\hat{c}^{a\dag}}_1\hat{c}^b_2\rangle,\\
&&\hbox{hole density:}\quad
D_{12}^{\alpha\,\beta}=\langle{\hat{D}^{\alpha\,\beta}}_{12}\rangle=\langle{\hat{d}^{\alpha\dag}}_1\hat{d}^\beta_2\rangle,\nonumber
\end{eqnarray}
\noindent where the indices $a,b,..$ label the conduction bands while
 $\alpha,\beta,...$ label the valence bands. The excitonic transition density $Y^{\alpha\,b}_{12}$ contributes to the interband transition polarization with the following term
\begin{equation}\label{polcrystal}
P= 2\,\hbox{Re}\,\left(\int_{r=r_1-r_2}d^3r\,
\sum_{cv}\,M_{21}^{cv*}Y_{12}^{vc}\right),
\end{equation}
and the diagonal elements (the matrices $C$ and $D$) correspond to
the densities of electrons
\begin{equation}\label{densityelectrons}
\rho_e=\left.-e\sum_c\,C^{cc}_{12}\right|_{r_1=r_2},
\end{equation}
\noindent and holes
\begin{equation}\label{densityholes}
\rho_h=\left.e\sum_v\,D^{vv}_{12}\right|_{r_1=r_2}.
\end{equation}
\noindent The matrices above are sub-matrices of the following density matrix
\begin{equation}
\underline{\underline{\hat{\rho}}}=\begin{pmatrix}
C_{cc'}&Y^*_{vc} \cr Y_{vc}&1-D_{vv'}\end{pmatrix}.
\end{equation}
\noindent The dynamics of the two-point matrices $Y,C,D$ is part
of the hierarchy of reduced density matrices, and is obtained from
the Heisenberg equations of motion
\begin{equation}\label{Liouville}
i\hbar\partial_t\underline{\underline{\hat{\rho}}}
=[\underline{\underline{\hat{\rho}}},{
H}]+i\hbar\partial_t\underline{\underline{\dot{\hat{\rho}}}}_{\rm
irrev},
\end{equation}
\noindent where the term
$\underline{\underline{\dot{\hat{\rho}}}}_{\rm irrev}$ describes
the irreversible dissipation and radiation decay processes due to all
dephasing processes. In this paper, we consider electron-electron interactions, electron-phonon interactions and the optical transitions caused by the electromagnetic field. In many practical calculations all irreversible processes are described in terms of two dephasing times $T_1, T_2$ which are taken as phenomenological constants that satisfy the following equation
\begin{eqnarray}
&&\left.\frac{\partial\underline{\underline{\hat{\rho}}}}{\partial
t}\right|_{\rm{irrev}}
=\\
&&\quad=-\begin{pmatrix}\frac{1}{T_1}[C(t)-C^{(0)}]&\frac{1}{T_2}[Y^*(t)-Y^{*(0)}]\cr
\frac{1}{T_2}[Y(t)-Y^{(0)}]&\frac{1}{T_1}[D(t)-D^{(0)}]
\end{pmatrix}.\nonumber
\end{eqnarray}
where the states with superscript ${(0)}$ denote the steady state solutions. The Hamiltonian in Eq. (\ref{Liouville}) can be written as
\begin{equation}\label{Htotal}
H = H_{ee} + H_{hh} + H_{eh} + H_{ph}
\end{equation}
where we have the electron-hole interaction
\begin{eqnarray}\label{Heh1}
&&H_{eh}=E_g-V_{12}+\frac{1}{2m_h}\left(p_1-eA_1\right)^2+\frac{1}{2m_e}\left(p_2+eA_2\right)^2\nonumber\\
&&+e\left(\Phi^h_1-\Phi^e_2\right),
\end{eqnarray}
the electron Hamiltonian
\begin{equation}
H_{ee}=\frac{1}{2m_e}\left[(p_2+eA_2)^2-(p_1-eA_1)^2\right]+e\left(\Phi^e_1-\Phi^e_2\right),
\end{equation}
and the hole Hamiltonian
\begin{equation}\label{Hhh1}
H_{hh}=\frac{1}{2m_h}\left[(p_2-eA_2)^2-(p_1+eA_1)^2\right]-e\left(\Phi^h_1-\Phi^h_2\right)
\end{equation}
where $E_g$ denotes the gap energy, $m_e, m_h$ are the effective masses of electrons and holes, respectively, $-V_{12}$ is the statically screened Coulomb potential, $M_0$ is the element in equation
(\ref{dipolematrixelement}) integrated over the real space, $A_j$ the vector potential of EM field at position ${\bf r}_j$, which may include an external magnetic field, $\Phi^{e/h}_j$ the
scalar, external or electromagnetically induced potential acting on electrons (or holes) at position ${\bf r}_j$, and $E_j$ denotes the electric field of the radiation at the point ${\bf r}_j$.  We
neglect for the moment the vectorial and tensorial indices, and
use the common notation for the momentum operators: $p_1=
-i\hbar\nabla_1$ etc. The electron-electron, hole-hole and
electron-hole exchange terms are included in the third term of the
Hamiltonian (\ref{2quanhamiltonian}), as given by expression
(\ref{HC}), and are obtained  by means of the random phase approximation (RPA) decoupling
scheme.~\cite{Stahl88}

The exciton-phonon interaction is given by
\begin{equation}\label{fonon_zaburzenie}
H_{ph}=a({\bf q})\,e^{-i\omega_{ph}t}V_p({\bf q},{\bf r})+\hbox{c.c.},
\end{equation}
with
\begin{eqnarray}\label{potencjal}
&&V_{p}({\bf q},{\bf r})=\sum\limits_{{\bf
R}_a}\left(\frac{\hbar}{2M\omega_{ph}}\right)^{1/2} e^{i\textbf{q}\textbf{R}_a}\textbf{e}_0\hbox{\boldmath$\nabla$}_r
V({\bf r}-{\bf R}_a).
\end{eqnarray}
where $\omega_{ph}$ is the angular phonon frequency and \textbf{q} is its wave vector. The $a({\bf q})$ is the phonon annihilation operator (with corresponding momentum $\hbar \textbf{q}$ and energy $\hbar \omega_{ph}$), its adjoint is the phonon creation operator, $V({\bf r}-{\bf R}_a)$ is the potential at the point ${\bf R_a}$, and ${\bf e}_0$ is the phonon polarization.

With above expressions, the Heisenberg equations (\ref{Liouville}) become a closed set of differential equations (``constitutive equations'') for $Y,C,D$, which can be obtained in explicit form using the following procedure:~
\cite{HuhnStahl84,StB87}\\
(i) setting up the Heisenberg equations of motion for the pair operators,\\
(ii) applying anti-commutation rules for the Fermion operators
$\hat{c}_j^{c\dag}, \hat{c}_j^c, \hat{d}_j^{v\dag}$ and
$\hat{d}_j^v$ to bring all operator products into normal order,\\
(iii) computing the expectation values of the relevant operators,\\
(iv) using an interpolation procedure to obtain  a continuum
dependence on the position variables (for example, Ref.\cite{StB87}),\\
(v) making use of the RPA to factorize four-point density matrices.\\
As a result we obtain the constitutive equations for the interband transition density amplitudes, which for any couple of bands are of the following form
\begin{eqnarray}\label{genconstitut}
&&-i\hbar
\partial_tY_{12}+H_{eh}Y_{12}+X^Y_{12}\\
&&={M_0}\left(E\delta_{12}-E_1C_{12}-E_2D_{21}\right)-
i\hbar\left(\frac{\partial Y_{12}}{\partial t}\right)_{\rm
irrev},\nonumber
\end{eqnarray}
\noindent with $X^Y_{12}$ the electron-hole exchange term, while the last dissipative term is responsible for irreversible excitation transfers. For intraband transitions (time dependence of the population of the band states) we obtain
\begin{eqnarray}
&&\label{CD}-i\hbar
\partial_tC_{12}+H_{ee}C_{12}+X^C_{12}\\
&&=-{M_0}\left(E_1Y_{12}-E_2Y^*_{21}\right)-
i\hbar\left(\frac{\partial C_{12}}{\partial t}\right)_{\rm irrev},\nonumber\\
&&\label{DV}
-i\hbar\partial_tD_{12}+H_{hh}D_{12}+X^D_{12}\\
&&=-{M_0}\left(Y_{21}E_1-Y^*_{12}E_2\right)-
i\hbar\left(\frac{\partial D_{12}}{\partial t}\right)_{\rm
irrev}\nonumber,
\end{eqnarray}
\noindent where $X^C_{12}$ and $X^D_{12}$ are the electron-electron and hole-hole
exchange terms, respectively. The
numerical subscripts are abbreviations for the coordinates, such as in $Y_{12}=Y({\bf r}_1,{\bf r}_2)$ etc. The exchange terms $X^Y, X^C, X^D$ read
\begin{eqnarray}\label{xy}
&&X^Y_{12}=\frac{i}{\hbar}\int
d^3r\,(V_{12}-V_{23})(Y_{13}C_{32}-D_{31}Y_{23}),\nonumber\\
&&X^C_{12}=\frac{i}{\hbar}\int\,(V_{13}-V_{23})(Y^*_{31}Y_{32}+C_{13}C_{32}),\\
&&X^D_{12}=\frac{i}{\hbar}\int\,(V_{13}-V_{23})(Y^*_{31}Y_{32}+D_{13}D_{32}),\nonumber
\end{eqnarray}
where $V_{12}$ is the statistically screened Coulomb potential.~\cite{Stahl90} It can be seen that
those terms represent 4-point correlations, but are not taken into account for 2-point correlation functions. It should be stressed that Hamiltonians are bilocal - they are defined in $({\bf r}_1,{\bf r}_2)$ space.

The electromagnetic fields  ${\bf A}$, ${\bf E}$ and $\Phi$ which appear in (\ref{Heh1})-(\ref{Hhh1}) are self-consistent fields that include the induced contributions produced by the sources
and those contained in $Y,C,D$. The spatial dependence of the functions $Y,C,D$ in the constitutive equations refers to a macroscopic scale. Microscopic structures can be taken into account by an appropriate choice of the parameters, as, for example, effective masses and $M_0$ in expressions (\ref{genconstitut}-\ref{DV}). The above expressions must be solved simultaneously with the Maxwell field equations
\begin{equation}\label{Maxwell}
-c^2\epsilon_0\nabla\times\nabla\times{\bf
E}-\epsilon_o\epsilon_b\ddot{\bf E}=\ddot{\bf P},
\end{equation}
\noindent where the polarization is given by expression (\ref{polcrystal}), and $c$ is the speed of light. A theoretical scheme containing higher order correlations has been presented in Ref.~\cite{Axt98}, which also discusses phonon-assisted transitions.

As mentioned above, the term $H_{ph}$ in the Hamiltonian (\ref{Htotal}) describes the electron-crystal lattice interaction. Using perturbation calculus, we obtain the matrix elements of
$H_{eh}$, under the condition of momentum conservation, and that of $H_{ph}$, which involves transfer of a specific momentum \textbf{q}. The calculated transition probability per unit time of a process in which the valence electron is scattered to the conduction state
$\psi_{ck_2}$, and a photon of energy $\hbar\omega$ with a phonon
of momentum $\textbf{q}=\textbf{k}_1-\textbf{k}_2$ and energy
$\hbar\omega_\textbf{q}$ are both absorbed, gives the absorption
coefficient in the case of the phonon absorption~\cite{bassani1975}
\begin{eqnarray}
&&\alpha_{ph, abs}(\omega)\\
&&=\left\{\begin{array}{ccc}0,&\hbox{for}&\hbar\omega<E_g-\hbar\omega_{ph}\\
C_1(\hbar\omega-E_g+\hbar\omega_{ph})^2n_{\hbox{\scriptsize\textbf{q}}},&\hbox{for}&\hbar\omega>E_g+\hbar\omega_{ph}\end{array}\right.,\nonumber
\end{eqnarray}
where $n_{\hbox{\scriptsize\textbf{q}}}$ is the phonon occupation number representing the number of available phonons with wave vector $\textbf{q}$, and $C_1$ is a constant. By considering all values of $\textbf{q}$ corresponding to energy $\hbar\omega_{ph}$, we use the distribution
\begin{equation}\label{phonons}
n_{ph}(\omega_{ph}) = \frac{1}{\exp\frac{\hbar\omega_{ph}}{k_B{\mathcal T}}-1},
\end{equation}
where $k_B$ is the Boltzmann constant, and $\mathcal T$ is the temperature.
The absorption coefficient in the case of phonon emission is given by
\begin{eqnarray}\label{alfaabs_nq}
&&\alpha_{ph, em}(\omega)\\
&&=\left\{\begin{array}{cc}0,&\hbar\omega<E_g+\hbar\omega_{ph}\\
C_1(\hbar\omega-E_g-\hbar\omega_{ph})^2n_{ph},&\hbar\omega>E_g+\hbar\omega_{ph}\end{array}\right.,\nonumber
\end{eqnarray}

The case of exciton formation in an indirect transition can be illustrated as follows. Let us consider
the case of parabolic non-degenerate energy bands with a maximum of the valence band at $\textbf{k}=0$, and a minimum of the conduction band at $\textbf{k}=\textbf{q}_0$, with
the dispersion
\begin{eqnarray}\label{ebands}
&&E_c(\textbf{k}_e)=\frac{\hbar^2(\textbf{k}_e-\textbf{q}_0)^2}{2m^*_e}+E_g,\nonumber\\
&&E_v(\textbf{k}_h)=-\frac{\hbar^2
\textbf{k}_h^2}{2m^*_h}.\end{eqnarray} For $k_{ex}=q_0$ we obtain
possible exciton energies
\begin{equation}
E_{ex}(q_0)=E_g-\vert E_{n\ell mH}\vert,
\end{equation} 
(see also equation (\ref{eigenvalues})) for states below the energy gap, and a
continuum for states above the indirect gap. The absorption coefficients due to indirect exciton transitions in a process in which a photon and a phonon are simultaneously absorbed, and for
the lowest exciton state, it is given by
\begin{eqnarray}\label{exphononabs}
&&\alpha_{ph, abs}(\omega)= 0\quad \hbox{for}\quad\hbar\omega<E_g- E_{1}-\hbar\omega_{ph},\\
&&\alpha_{ph, abs}(\omega)=C_2(\hbar\omega-E_g+E_1+\hbar\omega_{ph})^{\frac{1}{2}}n_{ph}\nonumber\\
&&\hbox{for}\;\hbar\omega>E_g- E_1-\hbar\omega_{ph}.\nonumber
\end{eqnarray}
where $C_2$ is a constant~\cite{bassani1975}, and we have written $E_1=\vert
E_{100H}\vert$. The analogous expression for the
absorption coefficient under the emission of a phonon has the form
\begin{eqnarray}\label{exphononemis}
&&\alpha_{ph, em}(\omega)= 0\quad \hbox{for}\quad \hbar\omega<E_g-E_{1}+\hbar\omega_{ph},\\
&&\alpha_{ph, em}(\omega)=C_2(\hbar\omega-E_g+E_1-\hbar\omega_{ph})^{\frac{1}{2}}n_{ph}\nonumber\\
&&\hbox{for}\;\hbar\omega>E_g- E_1+\hbar\omega_{ph}.\nonumber
\end{eqnarray} 
The total contribution of phonons to the absorption is then given by
\begin{equation}\label{phtotal}
\alpha_{ph,total}=\alpha_{ph, abs}(\omega)+\alpha_{ph,
em}(\omega).
\end{equation} 
It can be seen from the above expressions that the effect of phonons is relevant when
considering continuum states.

As discussed in Ref.~\cite{Valentin2008,Kim2015}, the phonon density of states contains two local maxima at 20~meV and 60~meV,  with a weighted average of approximately $\hbar\omega_{ph}=40$~meV. This simplified approach of taking an estimate of the average phonon energy provides a good fit to experimental data. In the same manner, as mentioned above, an integration of $n_{\textbf{q}}$ over all of values of $\textbf{q}$ yields an average phonon number $n_{ph}$ that can be used in equations (\ref{alfaabs_nq})-(\ref{exphononemis}).

\subsection{Two-photon absorption}
We next adapt the described RDMA approach to the case of two-photon absorption. We assume that static external fields are absent, thus neglecting the vector potential $A$ and the
scalar potential $\Phi^{e/h}$ in the Hamiltonian expressions (\ref{Heh1}-\ref{Hhh1}). In expressions (\ref{genconstitut}-\ref{DV}), the electromagnetic field $\bf E$ now includes two frequencies $\omega_a$ and $\omega_b$, and is written as
\begin{eqnarray}\label{efieldvacuum}
&&{\bf E}=\textbf{E}_{0a}\exp(
i\textbf{k}_{\textbf{a}}{\textbf{R}}-i\omega_a
t)\nonumber\\
&&+\textbf{E}_{0b}\exp(
i\textbf{k}_{\textbf{b}}{\textbf{R}}-i\omega_b t)+\hbox{c.c.},
\end{eqnarray}
where
\begin{equation}
\vert\textbf{k}_j\vert=\frac{\omega_j}{c}\sqrt{\epsilon(\omega_j)}=n_j\frac{\omega_j}{c},\quad
j=a,b,\end{equation} and $n_j$ are refractive indices at the
frequencies $\omega_j$, \textbf{R} is the electron-hole pair
center-of-mass coordinate
\begin{equation}\label{com}
{\bf R}={\bf R}_{12}=\frac{m_h{\bf r}_1+m_e{\bf r}_2}{m_h+m_e}.
\end{equation}
 The linear optical properties are calculated by solving the
interband equation (\ref{genconstitut}), supplemented by the
corresponding Maxwell equation, where the polarization
(\ref{polcrystal}) acts as a source. For computing the nonlinear
optical properties we use the entire set of constitutive equations
(\ref{genconstitut}-\ref{CD}). Although finding a general solution
of the equations is challenging, in special situations a solution
can be found. For example, if one assumes that the matrices $Y,C$
and $D$ can be expanded in powers of the electric field ${\bf E}$,
an iterative procedure can be used.

In general, solving for $Y$, $C$ and $D$ in the context of
two-photon absorption depends on the relation between the incoming
frequencies $\omega_a, \omega_b$ (and thus energies
$\hbar\omega_a,\;\hbar \omega_b$) and the fundamental gap energy
$E_g$, which enters as a parameter in the electron-hole
Hamiltonian. We consider two relevant cases, to be discussed
separately
\begin{enumerate}
\item When $\hbar\omega_a\;+\hbar \omega_b\,< 2E_g$, the
excitation of discrete excitonic states is possible. Therefore we
seek solutions in terms of eigenfunctions and eigenvalues of the
electron-hole Hamiltonian, taking also into account the phonons.
\item In the energy range $\hbar\omega_a\;+\hbar \omega_b\, >
2E_g$, we solve the equations (\ref{Ya}), (\ref{Ybezpot}) and
following equations for the matrices $C,D$ assuming
$V_{12}=0$~\cite{zeli_nonlinear_2019}, thus entering the energy
range represented by continuum states. The solution is obtained in
terms of an appropriate Green function.
\end{enumerate}


\subsection{Discrete states - linear susceptibility}\label{sec:discrete}
Our goal is to derive expressions for the NTA absorption coefficients. These can be obtained from the third-order nonlinear susceptibility, which, in turn, can be determined via an iterative procedure within the context of the RDMA. The first step in the
iteration consists of solving equation (\ref{genconstitut}), which at this stage takes on the form 
\begin{equation}\label{Ylin}
{i}\hbar\partial_tY^{(1)}-H_{eh}Y^{(1)}=-{\bf M}{\bf
E}+{i}\hbar\left(\frac{\partial Y^{(1)}}{\partial t}\right)_{{\rm
irrev}}.
\end{equation}
\noindent For the irreversible part we assume the simple form
\begin{equation}
\left(\frac{\partial Y^{(1)}}{\partial t}\right)_{{\rm
irrev}}=-\frac{1}{T_{2}} Y^{(1)}.
\end{equation}

\noindent In the discussion of nonlinear effects we also take  into account  the non-resonant parts of the  amplitude $Y$. The excitonic density $Y$ will consists of two parts, $Y_a, Y_b$, as defined by the angular frequencies $\omega_a$ and $\omega_b$. In addition, due to the valence band structure of the semiconducting material (Si) we must consider heavy-hole (H) and light hole (L) excitons. Considering optical transitions between the (H,L) valence bands and the conduction band, with the mentioned inclusion of both the resonant and anti-resonant parts, Eq. (12) generates eight equations: a pair for amplitude $Y_{aH}$: $Y_{aH-}^{(1)}\,\propto \exp(-{ i}\omega_a t)$ and for $Y^{(1)}_{aH+}\,\propto \exp({ i}\omega_a t)$

\begin{eqnarray}
& &{i}\hbar\left( i\omega_a+\frac{1}{T_{2}}\right)Y^{(1)}_{aH+}-
H_{eh}Y^{(1)}_{aH+}\nonumber\\
&&=-{\bf M}_H{\bf E}_a^*({\bf R},t),\\
\label{Ya}&&\nonumber\\
 & &{i}\hbar\left(-{
i}\omega_a+\frac{1}{T_{2}}\right)Y^{(1)}_{aH-}-
H_{eh}Y^{(1)}_{aH-}\\
&&=-{\bf M}_H{\bf E}_a({\bf R},t),\nonumber
\end{eqnarray}
with similar equations for $Y_{bH\pm}^{(1)}$, where $\textbf{M}_H$ is the transition dipole density.  Analogous equations hold for the amplitudes $Y_{a,bL\pm}^{(1)}$, with the appropriate transition dipole density $\textbf{M}_L$. In the following, we consider only one component of the vectors ${\bf E}$, ${\bf P}$ and ${\bf M}$, and focus our attention on the heavy hole exciton transition.

For the case of discrete exciton states, the exciton density in
the first step is found as
\begin{eqnarray}
&&Y^{(1)}_{daH-}=E_a({\bf R},t)\sum_{n\ell m}\frac{c_{n \ell mH}\varphi_{n}({\bf r})}{\hbar(\Omega_{n\ell mH}-\omega_a-{ i}/T_{2})},\nonumber\\
&&Y^{(1)}_{daH+}=E_a^*({\bf R},t)\sum_{n\ell m}\frac{c_{n\ell
m}\varphi_{n\ell mH}({\bf r})}{\hbar(\Omega_{n\ell m
H}+\omega_a-{i}/T_{2})},\label{Yd}
\end{eqnarray}
and similar expressions for $Y_{dbH\pm}^{(1)}$. The subscript
``d'' indicates the case of discrete excitonic states. The expansion coefficients are defined as follows
\begin{eqnarray}\label{basic}
&&c_{n\ell mH}=\int{ d}^3r M_H({\bf r})\varphi_{n\ell mH}({\bf r}),\nonumber\\
&&\hbar\Omega_{n\ell mH}=\hbar\Omega_{n\ell mH}=E_g+E_{n\ell mH}(\gamma_{aH}),\\
&&\varphi_{n\ell mH}=R_{n\ell H}(r)Y_{\ell
m}(\theta,\phi),\nonumber
\end{eqnarray}
$R_{n\ell H}$ are the hydrogen radial functions of the anisotropic
Schr\"{o}dinger equation~\cite{Zeit}, 
\begin{equation}
r=\sqrt{x^2+y^2+\frac{z^2}{\gamma_{aH}}},
\end{equation}
$Y_{\ell m}(\theta,\phi)$ are the spherical harmonics, and $E_{n\ell mH}$ the corresponding eigenvalues
\begin{eqnarray}\label{eigenvalues} &&
\phantom{nuc}E_{n\ell mH}=-\frac{\eta^2_{\ell
mH}(\gamma_{aH})R_H^*}{n^2},\quad
n=1,2,\ldots,\nonumber\\
&& \quad \ell=0,1,2,...n-1,\quad m=0,1,2,...\ell,
\end{eqnarray}
\noindent In the above, $R_H^*$ is the effective excitonic Rydberg energy
for the heavy hole exciton
\begin{equation}\label{excitonicRydberg}
R_H^*=\frac{\mu_{\parl
H}e^4}{2(4\pi\epsilon_0\sqrt{\epsilon_{\parl}\epsilon_z})^2\hbar^2},
\end{equation}
and the anisotropy parameter $\gamma_{aH}$ is defined as
\begin{eqnarray*}
&&\gamma_{aH}=\frac{\mu_{\parl
H}\epsilon_\parl}{\mu_{zH}\epsilon_z},
\end{eqnarray*}
where $\mu_{\parl H}$ and $\mu_{zH}$ are the heavy hole exciton reduced masses in the $x-y$ plane and in the $z$-direction, respectively,
 \begin{eqnarray}\label{masy1}
&&\frac{1}{\mu_{\parl H}}=\frac{1}{m_{e\parl}}+\frac{1}{m_{h\parl H}},\nonumber\\
&&\frac{1}{\mu_{zH}}=\frac{1}{m_{ez}}+\frac{1}{m_{hzH}}.
\end{eqnarray}
The anisotropic electron and heavy hole masses (in-plane and in
the $z$-direction) are listed in Table \ref{Sidata1}, $\epsilon_0$
is the vacuum dielectric constant, and $\epsilon_\parl,
\epsilon_z$ are relative dielectric tensor elements. The quantity $\eta_{\ell m H}(\gamma_{aH})$ is given by the following expression
\begin{eqnarray}\label{etaellm}
&&\eta_{\ell m H}(\gamma_{aH})=\int\limits_0^{2\pi}d\phi
\int\limits_0^{\pi} \frac{\vert Y_{\ell m} \vert^2 \sin \theta
d\theta }{\sqrt{\sin^2 \theta +\gamma_{aH}
\cos^2\theta}}.\label{etalm}
\end{eqnarray}

In what follows we consider only the excitonic $s$ states, with $\ell=m=0$, and denote
\begin{eqnarray}\label{varphi0}
&&\varphi_{n00H}=\varphi_{nH},\quad c_{n00H}=c_{nH},\nonumber\\
&&E_{n00H}=E_{nH},\quad
\hbar\Omega_{n00H}=\hbar\Omega_{nH}+E_n(\gamma_{aH}).
\end{eqnarray}
It should be noted that at room temperature, accounting for the relatively low binding energy of
excitons in Si (15 meV as in Ref.~\cite{Green_2013}), only the lowest excitonic state is relevant, so we set $n=1$.

The solutions for $Y_{da,b\pm}^{(1)}$ determined above permit the
calculation of the linear polarization
\begin{eqnarray}\label{chilin}
& &P_H^{(1)}\left(\omega\right)=\int  d^3{r}\left[Y_{daH-}^{(1)}+Y_{daH+}^{(1)*}\right]M_H^*({\bf r})\nonumber\\
&&+\int {d}^3r
\left[Y_{dbH-}^{(1)}+Y_{dbH+}^{(1)*}\right]M_H^*({\bf r})\nonumber\\
 &&=\frac{E_{0a}}{\hbar}\frac{2\vert
 c_{1H}\vert^2\Omega_{1H}}{\Omega^{2}_{1H
}-\left(\omega_a+{ i}/T_{2}\right)^2}\nonumber\\
&&+\frac{E_{0b}}{\hbar}\frac{2\vert
c_{1H}\vert^2\Omega_{1H}}{\Omega^{2}_{1H}-\left(\omega_b+{
i}/T_{2}\right)^2}\nonumber\\
&&=\epsilon_0\chi_{dH}^{(1)}(\omega_a)E_{0a}+\epsilon_0\chi_{dH}^{(1)}(\omega_b)E_{0b}.
\end{eqnarray}

The susceptibilities defined in equation (\ref{chilin}) can be
expressed in terms of the band parameters and, for energies below
the gap, when spatial dispersion is neglected, we obtain
\begin{eqnarray}
&&\chi^{(1)}_{dH}(\omega_j)=\epsilon_b
\frac{f_{1H}\Delta_{LTH}/R_H^*}{\left(E_{T1H}-\hbar\omega_j-
i\hbar/T_{2}\right)/R_H^*},
\end{eqnarray}
 where $j=a,b$, and $E_{T1H}$ is the energy of the first heavy hole exciton resonance. 
For the dipole density $M_H(\textbf{r})$ described by the formula
\begin{eqnarray}
&&M_H(\textbf{r})\\
&&=M_{0H}\,\frac{1}{\sqrt{4\pi}}\,\frac{1}{r\,r_{0H}^2\gamma_{aH}^{1/2}}e^{-r/r_{0H}}Y_{00}(\theta,\phi),\nonumber\end{eqnarray}
where $r_{0H}=(2\mu_{\parl H} E_g/\hbar^2)^{-1/2}$ is the
so-called coherence radius, 
the
anisotropy-dependent oscillator strength $f_{1H}$ has the form
\begin{equation}\label{anisotroscillstrength}
f_{1H}=\frac{\eta^3_{00H}(1-\eta_{00H}r_{0H}/a_H^*)}{(1+\eta_{00H}r_{0H}/a_H^*)^{4}},
\end{equation}
where $a_H^*$ is the heavy hole effective excitonic Bohr radius
\begin{equation}\label{excitonicBohr}
a_H^*=\frac{4\pi\hbar^2\epsilon_0\sqrt{\epsilon_{\parl}\epsilon_z}}{\mu_{\parl
H} e^2}.
\end{equation}
The longitudinal transverse splitting of the ground state is~\cite{RivistaGC}
\begin{equation}\label{LTsplit}
\frac{\Delta_{LT H}}{R^*}=2\frac{2\mu_{\parl H}}{\epsilon_0
\epsilon_b\pi\,a_H^*\hbar^2}M_{0H}^2,
\end{equation}
\noindent where the above mentioned bulk dielectric constant is given by $\epsilon_b=\sqrt{\epsilon_z\epsilon_\parl}$, and is $f_n\Delta_{LT H}$ for the excited states ($n>1$). Treating $\Delta_{LT H}$ as a known quantity, the above relation allows computation of the dipole matrix element $M_{0H}$.

\subsection{Discrete states - Nonlinear susceptibility}
To obtain the nonlinear response, the solutions for $Y_{daH,b\pm}^{(1)}$ are inserted as a source term in the conduction band equation (\ref{CD}) and the valence band equation (\ref{DV}). Note that each of these equations depends on the electromagnetic field. If the irreversible terms are well defined, the equations (\ref{CD},\ref{DV}) can be solved, and this second step of the iteration yields expressions for the density matrices $C_H$ and $D_H$. We use for the irreversible terms a linear relaxation time approximation
\begin{eqnarray}\label{irreversible}
&&\left(\frac{\partial C}{\partial t}\right)_{\rm
irrev}\\
&&=-\frac{1}{\tau}\left[C({\bf X}, {\bf r},t)-f_{0e}({\bf
r})C({\bf X}, {\bf r}=0,t)\right]- \frac{C(0)}{T_1},\nonumber
\end{eqnarray}
where ${\bf X}=({\bf r}_1+{\bf r}_2)/2$, $\tau$ is the
carrier relaxation time,  and $f_{0e}, f_{0h}$ are normalized
Boltzmann distributions for electrons and holes, respectively,
\begin{eqnarray}\label{Boltzmanne}
&&f_{0e}({\bf r})=\int d^3{q} f_{0e}({\bf q})e^{-i{\bf qr}}\\
&&= \exp\left(-\frac{m_{e\parl} k_B{\mathcal
T}}{2\hbar^2}\rho^2-\frac{m_{ez}\hbox{kT}}{2\hbar^2}z^2\right),\nonumber
\end{eqnarray}
The same type of expression holds for the
holes. The diagonal elements of the matrices are related to charge
densities (\ref{densityelectrons}, \ref{densityholes}), which are 
conserved quantities. Therefore, we assume that they relax to an
equilibrium normalized to the actual number of carriers. The 
relaxation time $T_1$ stands for the interband recombination time.~\cite{FrankStahl}
Using the irreversible terms (\ref{irreversible}) in the intraband
equations (\ref{CD}, \ref{DV}), and looking for stationary
solutions, we obtain the matrices $C,D$ in the form
\begin{eqnarray}\label{matricesCD}
&&C_H({\bf r})=-\frac{i}{\hbar}\left[\tau J_{CH}({\bf r})-\tau J_{CH}(0)+T_1f_{0e}({\bf r})J_{CH}(0)\right],\nonumber\\
&&D_H({\bf r})=-\frac{\rm i}{\hbar}\biggl[\tau J_{HH}({\bf r})\nonumber\\
&&-\tau J_{HH}(0)+T_{1H}f_{0hH}({\bf r})J_{HH}(0)\biggr],\nonumber
\end{eqnarray}
where
\begin{eqnarray}
&&J_{CH}=\frac{2{i}M_{0}\vert{E}_{0a}\vert^2}{\hbar}\biggl[\hbox{Im}\,g_H(-\omega_{a},
{\bf r})+ \,\hbox{Im}\,g_H(\omega_{a}, {\bf r})\biggr]\nonumber\\
&&+\frac{2{i}M_{0}\vert{E}_{0b}\vert^2}{\hbar}\biggl[\hbox{Im}\,g_H(-\omega_b,
{\bf r})+ \,\hbox{Im}\,g_H(\omega_b, {\bf r})\biggr],
\end{eqnarray}
and
\begin{equation} 
g_H(\pm\omega_j, {\bf r})=\sum_{n}\frac{c_{nH}\varphi_{nH}({\bf r})}{\Omega_{nH}\mp\omega_j-{i}/T_{2}},
\end{equation}
with $J_{CH}=J_{DH}$. These density matrices can then, in turn, be used as a source term for equation (\ref{genconstitut}), which can be solved to obtain expressions for $Y_{da,bH\pm}^{(3)}$ in the final step of the iteration.

The equations for the third-order coherent amplitudes
$Y^{(3)}_{jH\pm}$ now take on the form
\begin{eqnarray}
& &{i}\hbar\left( i\omega_a+\frac{1}{T_{2}}\right)Y^{(3)}_{aH+}-
H_{eh}Y^{(3)}_{aH+}\nonumber\\
&&=M_{0H}({E}^*C_H+{E}^*D_H)={E}^*({\bf R},t)\tilde{J}_H,\\
\label{Ya3}&&\nonumber\\
 & &{i}\hbar\left(-{
i}\omega_a+\frac{1}{T_{2}}\right)Y^{(3)}_{aH-}-
H_{eh}Y^{(3)}_{aH-}\\
&&=M_{0H}({E}^*C_H+{E}^*D_H)={E}^*({\bf R},t)\tilde{J}_H,\nonumber
\end{eqnarray}
with similar equations for $Y_{bH\pm}^{(1)}$,  where
\begin{eqnarray}
&
&\tilde{J}_H=-\frac{i}{\hbar}M_{0H}\left[T_{1}J_{CH}(0)f_{0eH}({\bf
r})\right.\nonumber\\
&&\left.+T_{1}J_{VH}(0)f_{0hH}(\textbf{r})\right].
\end{eqnarray}
\noindent Once $Y^{(3)}_{aH\pm}$ and $Y^{(3)}_{bH\pm}$ are know, the third-order polarization can be determined as 
\begin{eqnarray}\label{chilin3}
& &P_H^{(3)}\left(\omega\right)=\int  d^3{r}\left[Y_{aH-}^{(3)}+Y_{aH+}^{(3)*}\right]M_H^*({\bf r})\nonumber\\
&&+\int {d}^3r \left[Y_{bH-}^{(3)}+Y_{bH+}^{(3)*}\right]M^*_H({\bf
r}).
\end{eqnarray}
We consider nonlinear polarizations at the same frequency $\omega$ (or frequencies in the case 2PA) as that of an incident field, which means that we consider the susceptibilities $\chi^{(1)}$ and $\chi^{(3)}$ related to the excitonic amplitudes $Y^{(1)}$ and $Y^{(3)}$, respectively. 

Similar to the approach presented in Section \ref{sec_RDMA}, to account for the presence of phonons, we separate the polarization related to the emission of a phonon (subscript
``em'') and the absorption of a phonon (subscript ``abs''). The emission contribution has the form
\begin{eqnarray}\label{thirdorder}
&&P^{(3)}_{dH,em}(\omega_a,\omega_b)\\
&&=P^{(3)}_{daH,em}e^{-i\omega_at}+P^{(3)}_{dbH,em}e^{-i\omega_bt},\nonumber
\end{eqnarray}
where the polarization amplitudes are defined by
\begin{eqnarray}
&&P^{(3)}_{daH,em}=\epsilon_0\chi^{(3)}_{\hbox{\tiny d
H,self,em}}(\omega_a,\omega_a)\vert
E(\omega_a)\vert^2\,E(\omega_a)\nonumber\\
&&+\epsilon_0\chi^{(3)}_{\hbox{\tiny d
H,cross,em}}(\omega_a,\omega_b)\vert
E(\omega_b)\vert^2\,E(\omega_a),\nonumber\\
&&P^{(3)}_{dbH,em}=\epsilon_0\chi^{(3)}_{\hbox{\tiny d H,
self,em}}(\omega_b,\omega_b)\vert
E(\omega_b)\vert^2\,E(\omega_b)\nonumber\\
&&+\epsilon_0\chi^{(3)}_{\hbox{\tiny d H,
cross,em}}(\omega_b,\omega_a)\vert
E(\omega_a)\vert^2\,E(\omega_b),\nonumber
\end{eqnarray}
The nonlinear susceptibilities have the form
\begin{eqnarray}
&&\chi^{(3)}_{\hbox{ \tiny d H,self,em}}(\omega_j,\omega_j)\nonumber\\
&&=-(n_{ph}+1)\frac{2M_{0H}^2}{\epsilon_0}\frac{1}{T_2}\left(T_1+\frac{i\hbar\delta(\omega-2\omega_j)}{\hbar\omega+i\hbar/T_1}\right)\nonumber\\
&&\times\sum\limits_{\ell}\frac{c_{\ell H}(A_{\ell H}+B_{\ell
H})\hbar\Omega_{\ell H,em}}{(\hbar\Omega_{\ell
H,em})^2-(\hbar\omega_j+i\hbar T_2^{-1})^2}\\
&&\times\sum\limits_n\frac{c_{nH}\varphi_{nH}(0)}{(\hbar\Omega_{n H,em}-\hbar\omega_j)^2+(\hbar/T_2)^2},\label{chi3aa1}\nonumber\\
&&\chi^{(3)}_{\hbox{\tiny d H,cross,em}}(\omega_a,\omega_b)=-(n_{ph}+1)\frac{2M_{0H}^2}{\epsilon_0}\left(\frac{T_1}{T_2}\right)\nonumber\\
&&\times\sum\limits_{\ell}\frac{c_{\ell H}(A_{\ell H}+B_{\ell
H})\hbar\Omega_{\ell H,em}}{(\hbar\Omega_{\ell
H,em})^2-(\hbar\omega_a+i\hbar
T_2^{-1})^2}\\
&&\times\sum\limits_n\frac{c_{nH}\varphi_{nH}(0)}{(\hbar\Omega_{n
H,em}-\hbar\omega_b)^2+(\hbar/T_2)^2}.\nonumber
\end{eqnarray}

The additional cross term $\chi^{(3)}_{\hbox{\tiny
d H,cross,em}}(\omega_b,\omega_a)$ is obtained by permuting the
frequencies $\omega_a$ and $\omega_b$ in equation (\ref{thirdorder}).
The expressions $\hbar\Omega_{nH, em,abs}$ are the exciton resonance
energies that include the phonon energies $\hbar\omega_{ph}$
\begin{equation}\label{Omegaph}
 \hbar\Omega_{nH,em,abs}= E_g+E_{nH}\pm \hbar\omega_{ph},
\end{equation}
where ``+'' stands for phonon emission and ``-'' for phonon
absorption. 
The coefficients $A_{\ell H}, B_{\ell H}$ in equations (\ref{chi3aa1}) appearing in the expressions for
$Y_{da,bH\pm}^{(3)}$ have the form 
\begin{eqnarray}\label{Aell}
 &&A_{\ell H}=\int d^3 r\;\varphi_{\ell H}(\textbf{r})\;
 f_{0e}({\bf r})\nonumber\\
 &&=\int d^3 r\;\varphi_{\ell H}(\textbf{r})
 \exp\left(-\frac{\rho^2}{2\lambda_{th,e\parl}^2}-\frac{z^2}{2\lambda_{th,ez}^2}\right),\nonumber\\
 &&\\
 &&B_{\ell H}=\int d^3 r\,\varphi_{\ell H}(\textbf{r})\;f_{0hH}({\bf
 r})\nonumber\\
 &&\int d^3 r\;\varphi_{\ell H}(\textbf{r})
 \exp\left(-\frac{\rho^2}{2\lambda_{th,h\parl H}^2}-\frac{z^2}{2\lambda_{th,hzH}^2}\right),\nonumber
 \end{eqnarray}
 where $\lambda_{th,e},\lambda_{th,hH}$ are the so-called thermal
 lengths for electrons and holes, respectively,
 \begin{equation}\label{thermal}
 \lambda_{th,e}=\left(\frac{\hbar^2}{m_{e} k_B{\mathcal
 T}}\right)^{1/2},\;\;\lambda_{th,hH}=\left(\frac{\hbar^2}{m_{h} k_B{\mathcal
 T}}\right)^{1/2},\end{equation}
 determined for the appropriate masses ($\parl$ or $z$).
 The above
expressions are valid when
$\hbar\omega<E_g+\hbar\omega_{ph}$. 
For $\hbar\omega>E_g+\hbar\omega_{ph}$ one should replace $n_{ph}+1$ by
\begin{equation}
{\mathcal
C}(\hbar\omega-E_g-\hbar\omega_{ph})^2(n_{ph}+1).
\end{equation}
where $\mathcal C$ is a constant.~\cite{bassani1975}

Analogous expressions can be obtained for the susceptibilities
that include phonon absorption
\begin{eqnarray}
&&\chi^{(3)}_{\hbox{\tiny dH, self,abs}}(\omega_j,\omega_j)=
-n_{ph}\frac{2M_0^2}{\epsilon_0}\frac{1}{T_2}\left(T_1+\frac{i\hbar\delta(\omega-2\omega_j)}{\hbar\omega+i\hbar/T_1}\right)\nonumber\\
&&\times\sum\limits_{\ell}\frac{c_\ell(A_{\ell H}+B_{\ell
H})\hbar\Omega_{\ell H,abs}}{(\hbar\Omega_{\ell
H,abs})^2-(\hbar\omega_j+i\hbar
T_2^{-1})^2}\nonumber\\
&&\times\sum\limits_n\frac{c_{nH}\varphi_{nH}(0)}{(\hbar\Omega_{nH,abs}-\hbar\omega_j)^2+(\hbar/T_2)^2},\label{chi3aabs1}\\
&&\nonumber\\
&&\chi^{(3)}_{\hbox{\tiny dH, cross,abs}}(\omega_a,\omega_b)=-n_{ph}\frac{2M_{0H}^2}{\epsilon_0}\left(\frac{T_1}{T_2}\right)\nonumber\\
&&\times\sum\limits_{\ell}\frac{c_{\ell H}(A_{\ell
H}+B_\ell)\hbar\Omega_{\ell H,abs}}{(\hbar\Omega_{\ell
H,abs})^2-(\hbar\omega_a+i\hbar
T_2^{-1})^2}\nonumber\\
&&\times\sum\limits_n\frac{c_{nH}\varphi_{nH}(0)}{(\hbar\Omega_{nH,abs}-\hbar\omega_b)^2+(\hbar/T_2)^2},\label{chi3aabs2}
\end{eqnarray}
plus the additional cross term $\chi^{(3)}_{\hbox{\tiny dH,
cross,abs}}(\omega_b,\omega_a)$ obtained by permuting the input
frequencies in (\ref{chi3aabs2}). The above expressions are valid
when $$\hbar\omega<E_g-\hbar\omega_{ph}.$$ Otherwise, $n_{ph}$
should be replaced by
\begin{equation}
{\mathcal
C}(\hbar\omega-E_g+\hbar\omega_{ph})^2n_{ph}.
\end{equation} 
In section \ref{sec:2PA}, we use the expressions for the nonlinear
susceptibility above to determine the nonlinear absorption
coefficients in silicon.


\subsection{Continuum states}\label{sec:continuum}
In this section, we consider the case relevant to heavy-hole
exciton transitions. If $(\hbar\omega-\hbar\omega_{ph})>E_g$ for the case of
phonon emission, or $(\hbar\omega+\hbar\omega_{ph})>E_g$ for the
case of phonon absorption, then there is no generation of bound
exciton states, and continuum states constitute the final states
instead. In this case, $Y^{(1)}_H$ is calculated in the first
iteration step by setting $V_{12}=0$ in the electron-hole
Hamiltonian~\cite{zeli_nonlinear_2019}, giving rise to equations
of the form
\begin{eqnarray}\label{continuum}
&&\left(E_g \pm \hbar\omega \pm\hbar\omega_{ph}
-i{\frac{\hbar}{T_2}}-\frac{\hbar^2}{2\mu_{\parl
H}}\hbox{\boldmath$\nabla$}^2\right)Y_{H\pm}\nonumber\\
&&=\textbf{M}_H(\textbf{r})\textbf{E}.
\end{eqnarray}
Equation (\ref{continuum}) can be solved by means of the
appropriate Green function
\begin{equation}\label{Ybezpot}
Y_{H\pm}^{(1)}=\int d^3\, r'
g_{H\pm}(r,r')\textbf{M}_H(r',\theta,\phi)\textbf{E},\end{equation}
where
\begin{equation}
g_{H\pm}(r,r')=\frac{2\mu_{\parl
H}}{\hbar^2}\frac{\sinh\kappa_{H\pm}r^<}{4\pi\kappa_{H\pm}r^<r^>}e^{-\kappa_{H\pm}r^>},
\end{equation}
$r^<=\hbox{min}\,(r,r')$ and $r^>=\hbox{max}\,(r,r')$, and
\begin{equation}\kappa^2_{H\pm}=\frac{2\mu_{\parl H}}{\hbar^2}\left(E_g\pm\hbar\omega\pm\hbar\omega_{ph}-i\frac{\hbar}{T_2}\right).
\end{equation}
Assuming a linear polarization and the wave vector \textbf{E}
having a component $E_0$ in a direction $\alpha$, simultaneously
with the dipole density $\textbf{M}_H$ having a component $M_{0H}$
in the same direction and, for simplicity, using the dipole
density of the form 
\begin{equation}\label{Mdelta}
M_H(\textbf{r})=\frac{M_{0H}\delta(r-r_{0H})}{4\pi
r_{0H}^2},
\end{equation} 
we obtain
\begin{equation}\label{y2}
Y^{(1)}_{H\pm}=M_{0H}E_0\,g_{H\pm}(r,r_{0H}).
\end{equation} 
We may again define amplitudes of the form $Y^{(1)}_{caH\pm},
Y^{(1)}_{cbH\pm}$, where the subscript ``c'' now indicates the
involvement of continuum states. If, in the case of phonon
emission, we encounter $\kappa_H^2<0$, and introduce
\begin{equation}
\kappa_H=-i\tilde{\kappa_H},
\end{equation} 
where $$\tilde{\kappa}^2_{H-}=\frac{2\mu_{\parl H}}{\hbar^2}(\hbar\omega-\hbar\omega_{ph}-E_g+i\frac{\hbar}{T_2}),$$ and, in
this case, the Green function takes on the form
\begin{equation}
g_{H-}(r,r')=\frac{2\mu_{\parl
H}}{\hbar^2}\frac{\sin\tilde{\kappa}_{H\pm}r^<}{4\pi\tilde{\kappa}_{H\pm}r^<r^>}e^{i\tilde{\kappa}_{H\pm}r^>}.
\end{equation}
The linear terms for the case of phonon emission and absorption
and input frequency $\omega_a$ are found as
\begin{eqnarray}
&&Y^{(1)}_{caH-,em}(r)=M_{0H}E_{0a}g_{aH-,em}(r,r_{0H}),\nonumber\\
&&Y^{(1)}_{caH-,abs}(r)=M_{0H}E_{0a}g_{aH-,abs}(r,r_{0H}),
\end{eqnarray}
with
\begin{eqnarray}
&&g_{aH-,em}(r,r_{0H})=\frac{2\mu_{\parallel
H}}{\hbar^2}\frac{\sin(\tilde{\kappa}_{aH-,em}r^<)}{4\pi\tilde{\kappa}_{aH-,em}rr_0}e^{i\tilde{\kappa}_{aH-,em}r^>},\nonumber\\
&&\tilde{\kappa}_{aH-,em}^2=\frac{2\mu_{\parallel
H}}{\hbar^2}\left[\hbar\omega_a-(E_g+\hbar\omega_{ph})\right]+i\frac{2\mu_{\parallel
H}}{\hbar^2}\frac{\hbar}{T_2}.\nonumber
\end{eqnarray}

Similar expressions can be obtained for the amplitudes at input
frequencies $\omega_b$. The linear amplitudes thus obtained form
the source for calculating the $C_H$ and $D_H$ matrices, followed
by third step to determine $Y^{(3)}_{ca,bH\pm}$, similar to the
procedure described in section \ref{sec:discrete}. Once the
$Y^{(3)}_{ca,bH\pm}$ amplitudes are found, for both the phonon
emission and absorption process, we may write the nonlinear cross susceptibility for the continuum states as
\begin{equation}\label{chi3c}
\chi^{(3)}_{\hbox{\tiny cH,cross}}=\chi^{(3)}_{\hbox{\tiny
cH,cross,em}}(\omega_a,\omega_b)+\chi^{(3)}_{\hbox{\tiny
cH,cross,abs}}(\omega_a,\omega_b),
\end{equation}
with
\begin{eqnarray}
 &&\chi^{(3)}_{\hbox{\tiny
cH,cross,em}}(\omega_a,\omega_b)=\\
&&=-(n_{ph}+1)\frac{2}{\epsilon_0}2\frac{T_1}{\hbar}M_{0H}^4\left(\frac{2\mu_{\parallel
H}}{\hbar^2}\right)^2\nonumber\\
&&\times \frac{1}{4\pi
r_{0H}}\left(\frac{\sin(\tilde{\kappa}_{aH-,em}r_{0H})}{\tilde{\kappa}_{aH-,em}r_{0H}}\right)^2\nonumber\\
&&\times[\tilde{\kappa}_{bH-,em}r_{0H}+\tilde{\kappa}_{bH-,abs}r_{0H}]({\mathcal A}_{eH,em}+{\mathcal B}_{hH,em}),\nonumber\\
&&\chi^{(3)}_{\hbox{\tiny
cH,cross,abs}}(\omega_a,\omega_b)=\\
&&=-n_{ph}\frac{2}{\epsilon_0}2\frac{T_1}{\hbar}M_{0H}^4\left(\frac{2\mu_{\parallel
H}}{\hbar^2}\right)^2\nonumber\\
&&\times \frac{1}{4\pi
r_{0H}}\left(\frac{\sin(\tilde{\kappa}_{aH-,abs}r_{0H})}{\tilde{\kappa}_{aH-,abs}r_{0H}}\right)^2\nonumber\\
&&\times[\tilde{\kappa}_{bH-,em}r_{0H}+\tilde{\kappa}_{bH-,abs}r_{0H}]({\mathcal
A}_{eH,abs}+{\mathcal B}_{hH,abs}),\nonumber
\end{eqnarray}
where
\begin{eqnarray}
&&\tilde{\kappa}_{jH-,em}r_{0H}=\left(x_j-\frac{E_g+\hbar\omega_{ph}}{E_g}\right)^{1/2},\nonumber\\
&&\tilde{\kappa}_{j-,abs}r_{0H}=\left(x_j-\frac{E_g-\hbar\omega_{ph}}{E_g}\right)^{1/2},\nonumber\\
&&x_j=\frac{\hbar\omega_j}{E_g}.
\end{eqnarray} 
Here ${\mathcal A}_{eH,em}, {\mathcal B}_{hH,em}, {\mathcal
A}_{eH,abs}, {\mathcal B}_{hH,abs}$ are the continuum counterparts to expressions (\ref{Aell}), appropriate for discrete states. Making use of the relations (\ref{Aell}), we
obtain
\begin{eqnarray}
&&{\mathcal A}_{eH,em}={\mathcal A}_{eH,em}'+i{\mathcal
A}_{eH,em}'' =\frac{2\mu_{\parl H}}{\hbar^2}\frac{\sin
(\tilde{\kappa}_{aH-,em}r_{0H})}{\tilde{\kappa}_{aH-,em}r_{0H}}\nonumber\\
&&\times \int\limits_{\rho_{0H}}^\infty
\rho\,d\rho\int\limits_0^\infty dz
\frac{\exp(-i\tilde\kappa_{aH-,em})\sqrt{\rho^2+(z^2/\gamma_{aH})}}{\sqrt{\rho^2+(z^2/\gamma_{aH})}}
\nonumber\\
&&\times \exp\left(-\frac{\rho^2}{2\lambda_{th,e\parl}^2}-\frac{z^2}{2\lambda_{th,ez}^2}\right),
\end{eqnarray}
with an analogous expression for ${\mathcal A}_{eH,abs},\;{\mathcal B}_{hH,em}$, and ${\mathcal B}_{hH,abs}$. The value $\rho_{0H}=r_{0h}/a_H^*$.

\begin{small}
\begin{table}[ht!]
\caption{\small Band parameter values for Si, masses in free
electron mass $m_0$, H denotes heavy-, and L light-hole,
$R^*_{H,L}$ calculated from ($\mu_{\parl H,L}/\epsilon_b^2)\times
13600\,\hbox{meV}$,\, $a^*_{H,L}$ calculated from $(1/\mu_{\parl
H,L})\epsilon_b\times 0.0529\,\hbox{nm}$, $m_{e,dos}$
(density-of-state effective mass) calculated from
$6^{2/3}(m_{e\parl}m_{e\parl}m_{ez})^{1/3}$.}
\begin{center}
\begin{tabular}{p{.15\linewidth} p{.2\linewidth}p{.2\linewidth} p{.15\linewidth}  p{.2\linewidth}
} \hline
Parameter & Value (4.2\hbox{K}) &Value(300\hbox{K})&Unit&Reference\\
\hline $E_g$&1170 &1124 &meV& \cite{strehlow_compilation_2009}\\
$\Delta_{LTH}$&0.1&&{meV}&\\
$R_H^*$&15&16.56& meV& \\
$R^*_L$&7.94&9.6&meV&\\
$\gamma_1$&4.285&2.45&&\cite{Veni},\\
$\gamma_2$&0.339&0.194&&\cite{Veni},\\
$\gamma_3$&1.446&0.826&&\cite{Veni},\\
$m_{ez}$ & 0.9163& 1.09&$m_0$&\cite{palmer_semiconductors}\\
$m_{e\parl}$ & 0.1905&0.2& $m_0$&\cite{palmer_semiconductors}\\
$m_{e,dos}$&1.06&1.16&$m_0$&\\
$m_{hzH}$&0.28&0.485&$m_0$&Eq.(\ref{masyefektqw}),\cite{singh_physics_1993}\\
$m_{h\parl H}$ & 0.72& 1.25&$m_0$&Eq. (\ref{masyefektqw})\\
$m_{hzL}$&0.2&0.35&$m_0$&\\
$m_{h\parl L}$&0.14&0.24&$m_0$&\\
$\mu_{zH}$  &0.214 &0.336&$m_0$&\\
$\mu_{\parl H}$&0.15&0.172 &$m_0$&\\
$\mu_{zL}$&0.164&0.26&$m_0$&\\
$\mu_{\parl L}$&0.08&0.1&$m_0$&\\
$a^*_H$&4.13&3.66&nm&\\
$a^*_L$&7.74&6.3&nm&\\
$r_{0H}$&0.46&0.44& nm&\\
$r_{0L}$&0.64&0.58&nm&\\
$\epsilon_b$&11.7 &11.9&&\\
$T_2$ & 0.1 && ns & \cite{Green2013}\\
$T_1$ & 8 && ns & fitting\\
$\hbar\omega_{ph}$ & 40 && meV & fitting, \cite{Kim2015}\\
\end{tabular} \label{Sidata1}\end{center}
\end{table}
\end{small}
\begin{small}
\begin{table}[ht!]
\caption{\small Anisotropy parameters for Si, excitonic energies
calculated from Eq. (\ref{eigenvalues})}.
\begin{center}
\begin{tabular}{p{.2\linewidth} p{.2\linewidth}p{.2\linewidth} p{.15\linewidth}p{.15\linewidth}
} \hline
Parameter & Value (4.2$\,$K) &Value (300$\,$K)&Unit&Reference\\
\hline $\gamma_{aH}$&0.7&0.51& &Eq.(\ref{etalm})\\
$\eta_{00H}$&1.058&1.1&&Eq.(\ref{etalm})\\
$\vert E_{10H}\vert$&16.79&20&meV&\\\hline
\end{tabular} \label{Sidata2}\end{center}
\end{table}
\end{small}

\subsection{Nonlinear absorption coefficients}\label{sec:2PA}
The propagation of the field components $E_{0a}$ and $E_{0b}$ in
the semiconductor follow from the wave equation with $P^{(3)}$ as
a source term, which, after making the well-known slowly varying
amplitude approximation, yield the following coupled equations for
the field amplitudes
\begin{eqnarray}\label{wave,a}
&&\frac{\partial E_{0a}}{\partial
z}=i\frac{\omega_a}{n_a\,c}\chi^{(3)}_{\hbox{\tiny
self H}}(\omega_a,\omega_a)\vert E_{0a}\vert^2\,E_{0a}\\
&&+i\frac{\omega_a}{n_a\,c}\chi^{(3)}_{\hbox{\tiny cross
H}}(\omega_a,\omega_b)\vert E_{0b}\vert^2\,E_{0a}\;,\nonumber
\end{eqnarray}
\begin{eqnarray}\label{wave,b}
&&\frac{\partial E_{0b}}{\partial
z}=i\frac{\omega_b}{n_b\,c}\chi^{(3)}_{\hbox{\tiny
self H}}(\omega_b,\omega_b)\vert E_{0b}\vert^2\,E_{0b}\\
&&+i\frac{\omega_b}{n_b\,c}\chi^{(3)}_{\hbox{\tiny cross
H}}(\omega_b,\omega_a)\vert E_{0a}\vert^2\,E_{0b}\;.\nonumber
\end{eqnarray}
From the equation above, the intensities of the input beams can be
found as
\begin{eqnarray}\label{Iaa}
&&I_a=2\epsilon_0\,n_a\,c\,\vert E_{0a}\vert^2,\nonumber\\
&&\frac{\partial I_a}{\partial
z}=2\epsilon_0\,n_a\,c\,\left[E_{0a}^*\frac{\partial
E_{0a}}{\partial z}+E_{0a}\frac{\partial E^*_{0a}}{\partial
z}\right].
\end{eqnarray}
Similar expressions are obtained for $I_b$, resulting in the
following set of coupled equations (\ref{wave,a})-(\ref{wave,b})
we obtain the set of equations
\begin{eqnarray}\label{2PAcoeff}
&&\frac{\partial I_a}{\partial
z}=-\alpha_2(\omega_a,\omega_a)\,I_a^2(z)
-\alpha_2(\omega_a,\omega_b)\,I_a(z)I_b(z),\nonumber\\
&&\,\\
&& \frac{\partial I_b}{\partial
z}=-\alpha_2(\omega_b,\omega_b)\,I_b^2(z)
-\alpha_2(\omega_b,\omega_a)\,I_b(z)I_a(z),\nonumber
\end{eqnarray}

These equations define the nonlinear absorption coefficients $\alpha_2$ as 
\begin{eqnarray}\label{coefficients2PA}
&&\alpha_{2H}(\omega_a,\omega_a)=\frac{\omega_a}{2\epsilon_0\,n_a^2c^2}\,\chi^{(3),I}_{\hbox{\tiny
self H}}(\omega_a,\omega_a),\nonumber\\
&&\alpha_{2H}(\omega_a,\omega_b)=\frac{\omega_a}{2\epsilon_0\,n_a\,n_b\,c^2}\,\,\chi^{(3),I}_{\hbox{\tiny
cross H}}(\omega_a,\omega_b),\nonumber\\
&&\\
&&\alpha_{2H}(\omega_b,\omega_a)=\frac{\omega_b}{2\epsilon_0\,n_a\,n_b\,c^2}\,\chi^{(3),I}_{\hbox{\tiny
cross H}}(\omega_b,\omega_a),\nonumber\\
&&\alpha_{2H}(\omega_b,\omega_b)=\frac{\omega_b}{2\epsilon_0\,n_b^2c^2}\,\,\chi^{(3),I}_{\hbox{\tiny
self H}}(\omega_b,\omega_b),\nonumber
\end{eqnarray}
where $\chi_H^{(3),I}$ indicates the imaginary part of
$\chi_H^{(3)}$. The refractive indices
$n_a=n(\omega_a),n_b=n(\omega_b)$ are defined by the real parts of
the linear susceptibility $\chi^{(1)}$, defined as
\begin{eqnarray}\label{refractive}
&&n_a=\\
&&\left\{\left[\epsilon_b\left(1+\frac{f_1\Delta_{LT}[\hbar\Omega_{n,em}/E_g-x_a]}{[(\hbar\Omega_{n,em}/E_g)-x_a]^2+\gamma_2^2}\right)\right]\right\}^{1/2},\nonumber
\end{eqnarray}
with analogous formulas for $\omega_b$ and the absorptive term
$\hbar\Omega_{n,abs}$. Here, $\gamma_2=\hbar/(E_gT_2)$, $f_1$ is the oscillator strength
(here for the heavy hole exciton), written as
\begin{equation}
f_{1H}=\frac{\eta_{00H}^3}{[1+\eta_{00H}(r_{0H}/a^*_H)]^3},
\end{equation}
see Eq. (\ref{anisotroscillstrength}).

The susceptibilities and, consequently, the nonlinear absorption
coefficients, are composed of four components: two corresponding to
the contributions of discrete and continuous states, and each of
them containing terms related to phonon emission and absorption
\begin{equation}
\alpha_{2H}(\omega_a,\omega_b)=\alpha_{2dH}(\omega_a,\omega_b)+\alpha_{2cH}(\omega_a,\omega_b),\label{alpha2_d_c}\end{equation}
For the discrete states, we only consider the lowest exciton state
with $n=1$, yielding the following form
\begin{eqnarray}\label{crossnew}
&&\alpha_{2dH}(\omega_a,\omega_b)\nonumber\\
&&=-2
{x_a\gamma_2\alpha'}\left(\frac{T_1}{T_2}\right)\biggl\{\frac{\epsilon_b}{n_an_b}\frac{n_{ph}+1}{(\hbar\Omega_{1H,em}/E_g-x_b)^2+\gamma_2^2}\nonumber\\
&&\times\frac{\hbar\Omega_{1H,em}/E_g}{[(\hbar\Omega_{1H,em}/E_g)^2-x_a^2]^2+(2\gamma_2x_a)^2}\nonumber\\
&&+\frac{\epsilon_b}{n_an_b}\frac{n_{ph}}{(\hbar\Omega_{1H,abs}/E_g-x_a)^2+\gamma_2^2}\nonumber\\
&&\times\frac{\hbar\Omega_{1H,abs}/E_g}{[(\hbar\Omega_{1H,abs}/E_g)^2-x_b^2]^2+(2\gamma_2x_b)^2}\biggr\},
\end{eqnarray}
where the constant $\alpha'$ is defined as
\begin{eqnarray*}
&&\alpha_H'(\omega_a,\omega_b)\\
&&=2\frac{4E_g}{\epsilon_0\hbar
n_a\,n_b2c^2E_g^3}\frac{2M_{0H}^2}{\epsilon_0}\left[\varphi_{1H}(0)A_{1H}+\varphi_{1H}(0)B_{1H}\right]c_{1H}^2\\
&&=2\frac{1}{4}\epsilon_0^2\epsilon_b^2\pi^2a_H^{*6}\Delta_{LTH}^2\times\frac{4}{\pi\epsilon_0\hbar
n_a\,n_bc^2E_g^2}\\
&&\times\left[\varphi_{1H}(0)A_{1H}+\varphi_{1H}(0)B_{1H}\right]\left(\frac{\eta_{00H}}{a^*_H}\right)^3\frac{1}{(1+\eta_{00H}r_{0H}/a^*_H)^2}\\
&&=2\frac{\pi\epsilon_b^2\Delta_{LTH}^2(\eta_{00H}a^*_H)^3}{E_g^2\hbar
c^2\,n_a\,n_b}\frac{\left[\varphi_{1H}(0)A_{1H}+\varphi_{1H}(0)B_{1H}\right]}{(1+\eta_{00H}r_{0H}/a^*_H)^2}.
\end{eqnarray*}

The continuum states contribution has the form
\begin{equation}
\alpha_{2cH}(\omega_a,\omega_b)=\alpha_{2cH,em}(\omega_a,\omega_b)+\alpha_{2cH,
abs}(\omega_a,\omega_b),\label{alpha_c}
\end{equation} 
where
\begin{eqnarray}
&&\alpha_{2cH,em}(\omega_a,\omega_b)\nonumber\\
&&=-\frac{4E_g(n_{ph}+1)}{\epsilon_0\hbar\,n_a\,n_b\,c^2}M_{0H}^4\left(\frac{2\mu_{\parallel
H}}{\hbar^2}\right)^2\frac{1}{4\pi
r_{0H}}\frac{T_1}{\hbar}\nonumber\\
&&\times
x_a\left(\frac{\sin(\tilde{\kappa}_{aH-,em}r_{0H})}{\tilde{\kappa}_{aH-,em}r_{0H}}\right)^2[\tilde{\kappa}_{bH-,em}r_{0H}+\tilde{\kappa}_{bH-,abs}r_{0H}]\nonumber\\
&&\times \left({{\mathcal A}''_{eH,em}+{\mathcal
B}''_{hH,em}}\right),
\end{eqnarray}
and 
\begin{eqnarray}
&&\alpha_{{2cH,abs}} (\omega_a,\omega_b)\nonumber\\
&&=-\frac{4E_g\,n_{ph}}{\epsilon_0\hbar\,n_a\,n_b\,c^2}M_{0H}^4\left(\frac{2\mu_{\parallel
H}}{\hbar^2}\right)^2\frac{1}{4\pi
r_{0H}}\frac{T_1}{\hbar}\nonumber\\
&&\times x_a\left(\frac{\sin(\tilde{\kappa}_{aH-,abs}r_{0H})}
{\tilde{\kappa}_{aH-,abs}r_{0H}}\right)^2[\tilde{\kappa}_{bH-,em}r_{0H}+\tilde{\kappa}_{bH-,abs}r_{0H}]\nonumber\\
 &&\times \left({{\mathcal A}''_{eH,abs}+{\mathcal B}''_{hH,abs}}\right).
\end{eqnarray}
The solutions obtained in the two regimes of discrete and
continuous states are smoothly connected via the use of
hyperbolic tangent functions. Using the above formula, we have
calculated the 2PA coefficient $\alpha_2(\omega_a,\omega_b)$ as a
function of the energies $\hbar\omega_a+\hbar\omega_b$. The band
parameters used in the calculations are listed in Tables
\ref{Sidata1} - \ref{Sidata3}.


The masses are calculated from Luttinger parameters
\begin{eqnarray}\label{masyefektqw}
m_{hzH}&=&\frac{m_0}{\gamma_1-2\gamma_2},\nonumber\\
m_{h\parl H}&=&\frac{m_0}{\gamma_1-2\gamma_3},\nonumber\\
m_{hzL}&=&\frac{m_0}{\gamma_1+2\gamma_2},\\
m_{h\parl L}&=&\frac{m_0}{\gamma_1+2\gamma_3}.\nonumber
\end{eqnarray}
\noindent With the reduced electron and hole masses, the thermal lengths are given by
\begin{eqnarray}\label{lambdath}
&&\overline{\lambda}_{th,e}=\left(\frac{\hbar^2}{m_{e\parl}
k_B{\mathcal
 T}}\right)^{1/2}=\left(2\times\frac{\hbar^2}{2\mu_{\parl H}}\times\frac{\mu_{\parl H}}{m_{e\parl}}\frac{1}{k_B{\mathcal
 T}}\right)^{1/2}\nonumber\\
 &&=\left(2\times\frac{\mu_{\parl H}}{m_{e\parl}}\frac{R^*_H}{k_B{\mathcal
 T}}\right)^{1/2},\nonumber\\
 &&\overline{\lambda}_{th,hH}=\left(\frac{\hbar^2}{m_{h\parl} k_B{\mathcal
 T}}\right)^{1/2}\nonumber\\
 &&k_BT =8.617\times 10^{-2}\;\frac{\hbox{meV}}{\hbox{K}}.
\end{eqnarray}
The calculated values are summarized in Table \ref{Sidata3}.
\begin{small}
\begin{table}[ht!]
\caption{\small Thermal electron and hole lengths for Si, and
expressions $\varphi_{1H}(0)A_{1H}\;, \varphi_{1H}(0)B_{1H}$}
\begin{center}
\begin{tabular}{p{.2\linewidth} p{.2\linewidth}p{.2\linewidth} p{.2\linewidth}
} \hline
Quantity & $\hbox{Value 4.2K}$ &Value 300$\,$K&Reference\\
\hline $\overline{\lambda}_{th,e\parl}$&8.08&1.05& Eq. (\ref{lambdath}) \\
$\overline{\lambda}_{th,ez}$&3.68&0.45& Eq. (\ref{lambdath})\\
$\overline{\lambda}_{th,h\parl H}$&4.15&0.42&Eq. (\ref{lambdath})\\
$\overline{\lambda}_{th,hz H}$&6.66&0.67&Eq. (\ref{lambdath})\\
$\varphi_{1H}(0)A_{1H}$&$4\times 1.52$&$4\times 0.173$&\\
$\varphi_{1H}(0)B_{1H}$&$4\times 1.34$&$4\times 0.06$&\\\hline
\end{tabular} \label{Sidata3}\end{center}
\end{table}
\end{small}

\section{Experimental}
We perform optical experiments to determine nonlinear absorption coefficients of silicon to supplement existing data in the literature, spanning a wider range of $x=\frac{\hbar(\omega_a+\omega_b)}{2E_g}$ values. For this purpose, we perform cross correlation experiments, using optical pulses derived from a 1-kHz amplified femtosecond laser system (Spitfire Ace, Spectra Physics). The laser seeds two optical parametric amplifiers (OPA, TOPAS-Prime, Light Conversion), where one OPA is used as a source of near-infrared (NIR) probe radiation between 1150~nm - 1350~nm (1.08~eV - 0.91~eV), producing pulses in the range of 100-150~fs. The signal and idler pulses from the second OPA system are used to generate MIR pump pulses through the process of difference frequency generation (DFG) in the 2480~nm to 4651~nm (0.5~eV - 0.27~eV) range, producing 175-265~fs pulses. The MIR pump is modulated using an optical chopper at a frequency of 500~Hz that is synchronized to the laser output (MC2000B, Thorlabs).

We use a 280~$\mu$m thick [100] silicon window (99.999\% purity, University Wafer) as the sample target. The MIR and NIR pulses are focused in a non-collinear arrangement on the Si target, using normal incidence for the pump beam and a $\sim20$ degree incidence angle for the probe beam. Temporal overlap is controlled through an automated translation stage (GTS150, Newport) in the probe arm, producing a cross-correlation of pump induced probe absorption via NTA. The remaining probe is attenuated by an OD = 3 neutral density filter and detected using a home-built InGaAs photodiode. The modulated change induced by the MIR pump is analyzed by a lock-in amplifier (SR860, Stanford Research Systems). The resulting cross-correlation is then used to extract the nonlinear absorption coefficient $\alpha_2$ as described by Negres et al.~\cite{negres_experiment_2002}

\begin{figure}[H]
\includegraphics[width=0.45\textwidth]{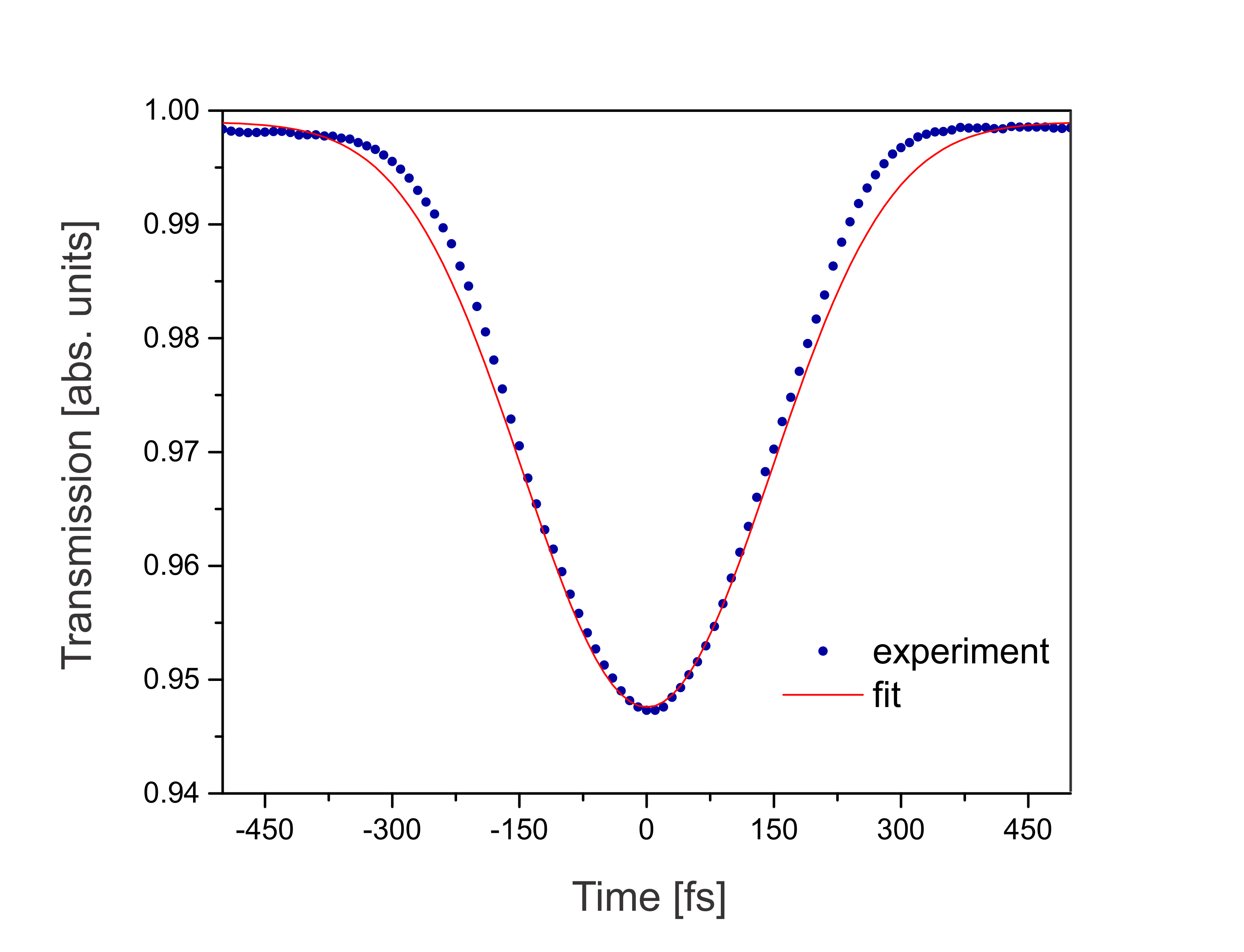}
\caption{\label{fig:a2fit} Exemplary pump-probe data and corresponding nonlinear transmittance fit. The photon energies used here are 0.91 eV and 0.376 eV.}
\end{figure}

Figure \ref{fig:a2fit} shows a representative cross correlation and its corresponding fit, and the tabulated data of extracted $\alpha_2$ values are presented in Table \ref{table:t1}. A full description of the data analysis is presented in the Supplemental Materials. Pump-to-probe beam radius ratios were maintained to at least 8:1 as measured by knife edge scan. The low irradiance of the probe beam in conjunction with the use of a lock-in amplifier ensures that any degenerate two-photon absorption present is excluded from the measured signal. The pump beam is always the longer wavelength beam, minimizing free carrier absorption induced by the probe via three- or four-photon absorption.

\begin{table}
\begin{center}
\caption{Experimentally obtained values for $\alpha_2$ in cm/GW}
\begin{tabular}{| c | c | c | c | c |} 
 \hline
 Wavelength (nm) & 2480 & 3295 & 3755 & 4220 \\ 
 \hline\hline
 1150 & 1.099 & 0.71 & 0.823 & 0.955 \\ 
 \hline
 1200 & 0.951 & 0.638 & 0.705 & 0.817 \\
 \hline
 1250 & 0.846 & 0.564 & 0.501 & 0.481 \\
 \hline
 1350 & 0.634 & 0.452 & 0.385 & 0.307 \\
 \hline
\end{tabular}

\label{table:t1}
\end{center}
\end{table}

\section{Results and Discussion}
\begin{figure}[H]
\includegraphics[width=0.5\textwidth]{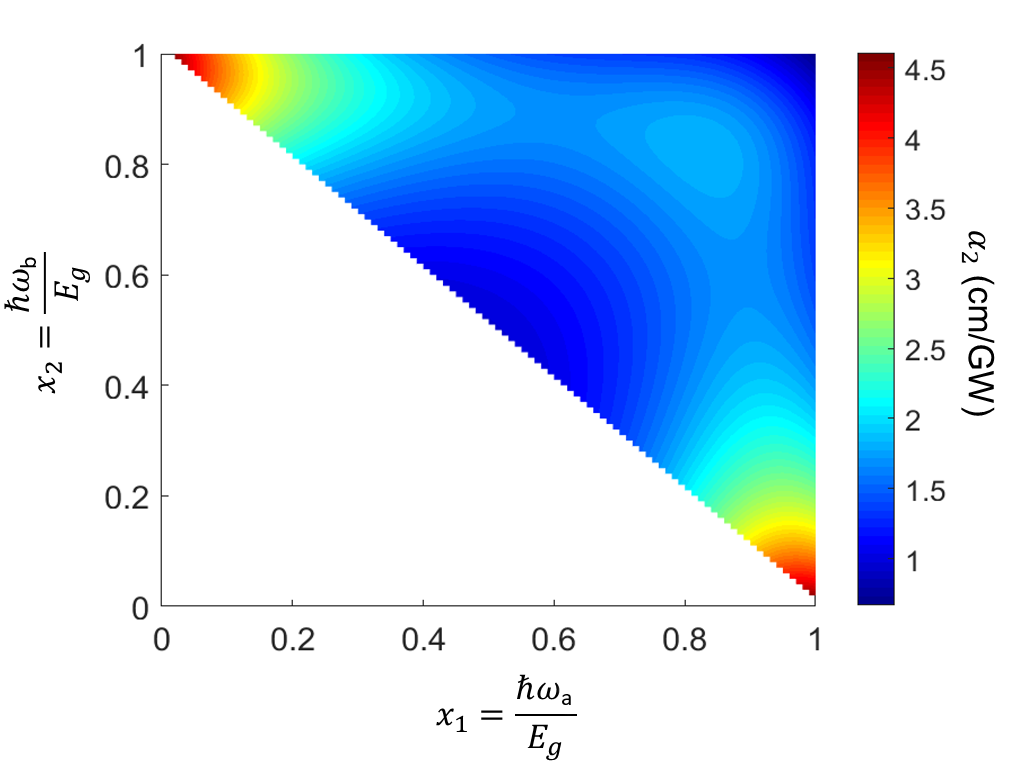}
\caption{\label{fig:triscale} Scaling behavior of the two photon absorption coefficient $\alpha_2$ (Eq. 103) as a function of normalized photon energy while both photons are below the gap.}
\end{figure}

The derived form of $\alpha_{2H}$ that follows from the RDMA contains contributions from the two regimes of exciton production, namely the contributions from discrete states and
continuum states, see equation (\ref{alpha2_d_c}). The discrete regime is visually depicted
in Figure \ref{fig:triscale}, which highlights the two-dimensional
dispersion of $\alpha_2$. The analysis recognizes the presence of
a single discrete excitonic state below the band edge,
corresponding to the creation of a bound state through absorption
of a phonon, $\hbar\Omega_{abs}$, which may provide considerable
enhancement to the 2PA process. This is manifested by the
presence of resonant denominators for the discrete regime, which
can be written in simplified form from equation (\ref{crossnew})
as 
\begin{eqnarray}
    \alpha_{2dH}(\omega_a,\omega_b)\propto \frac{1}{(\hbar\Omega_{abs}/E_g-\hbar\omega_b/E_g)^2+\gamma_2^2}\times\nonumber\\
    \frac{\hbar\Omega_{em}/E_g}{[(\hbar\Omega_{em}/E_g)^2 - (\hbar\omega_a)^2]^2+(2\gamma_2\hbar\omega_a)^2}
\end{eqnarray}
This single state is the only one considered due to the low
binding energy of Si, which precludes the formation of bound
excitons with higher principal quantum numbers. When
$\hbar\omega_b/\hbar\omega_a$ is detuned away from degeneracy, we
observe an increase of $\alpha_2$ that is as much as $5\times$ the
equivalent degenerate response, amplifying the process as it
becomes doubly resonant with the lowest discrete state and the
edge of the continuum at room temperature. This behavior is
evident in the corners of Figure \ref{fig:triscale}. While smaller
in magnitude, the $\hbar\Omega_{abs}$ state provides additional
enhancement when at least one incident photon approaches this
energy. An alternate resonance condition exists when both photons
are resonant with the discrete exciton level at approximately 95\%
of the band gap. This is a singly resonant process where the
intermediate state is the lowest excitonic state, providing some
enhancement to the two-photon process.

If the the energy of the incident photons exceeds the bandgap
energy, then the two-photon absorption process proceeds entirely
via continuum states. This can be accessed by either phonon
absorption or emission, and the nonlinear absorption coefficient
has the form as in equation (\ref{alpha_c}). Beyond the band edge,
the continuum causes $\alpha_{2cH}$ to further increase up to the
point where the energy of at least one of the photons has enough
energy to reach the direct gap of silicon at $E_g=3.43~\rm{eV}$.
This increase is due to the oscillatory behavior of the linear
coherent amplitude of the exciton density $Y^{(1)}$, which
contributes to the induced polarization of the medium in order for
two-photon absorption to occur.

\begin{figure}[H]
\includegraphics[width=0.5\textwidth]{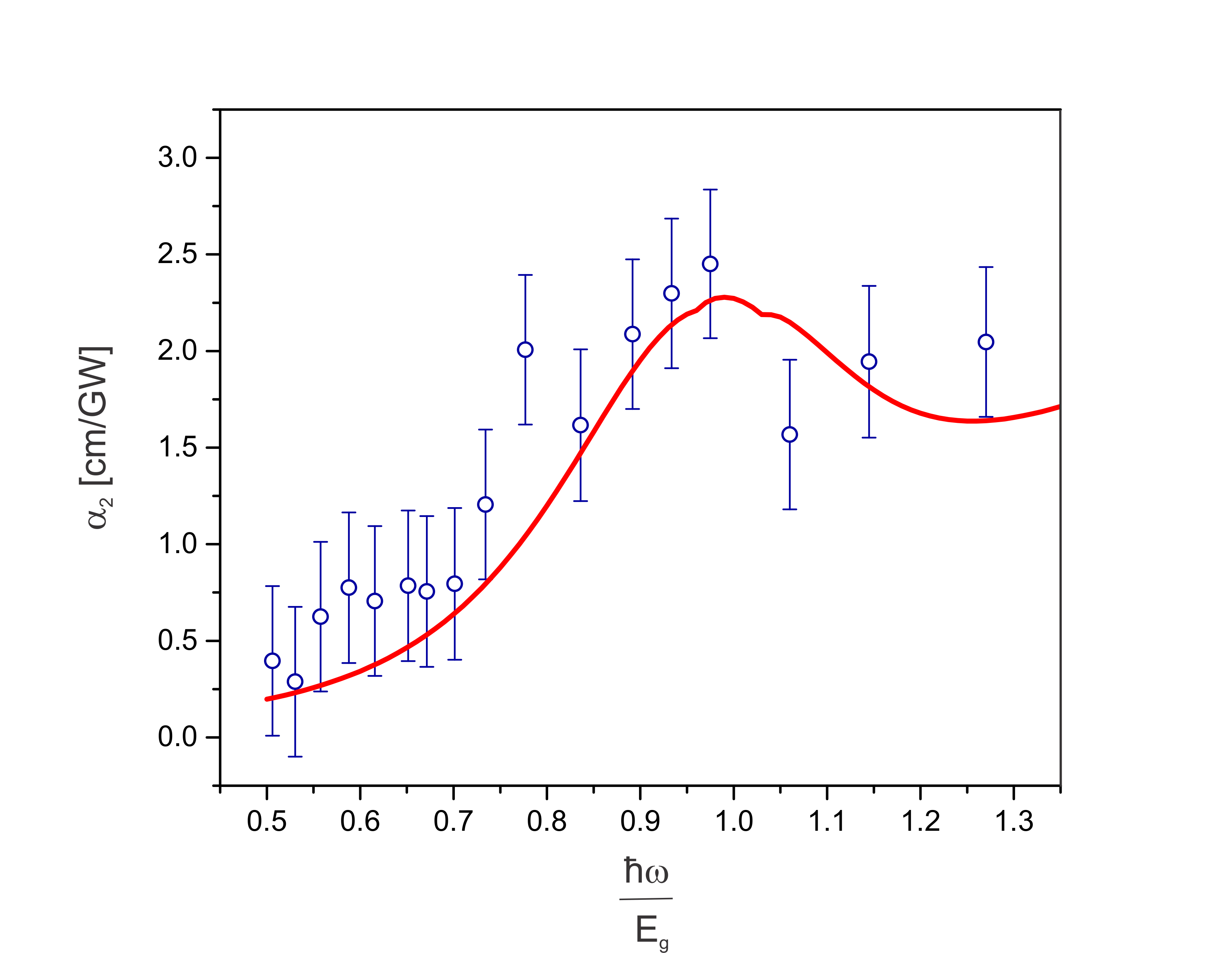}
\caption{\label{fig:bristow} Scaling of the degenerate two-photon absorption coefficient $\alpha_2$ as a function of $x={\hbar\omega}/{E_g}$. Squares are obtained from Ref~\cite{bristow_two_2007} and thick solid line results from the RDMA analysis of two-photon absorption.}
\end{figure}

To validate the performance of the model, we first apply it to
explain an experimental data set of the degenerate two-photon
(DTA) cross section obtained from an open aperture Z-scan reported
in Ref~\cite{bristow_two_2007}. As can be seen in Figure
\ref{fig:bristow}(a), $\alpha_2$ for the DTA process reveals a
resonance-like behavior as a function of photon energy. A previous
analysis using a model for direct transitions, corrected for the
center of mass energy when phonon scattering is
involved~\cite{Garcia2006}, reproduced the general dependence of
$\alpha_2$ on the photon energy, but it failed to predict the
observed resonance structure. The RDMA approach (indicated by the
solid line in Figure \ref{fig:bristow}), on the other hand,
predicts a resonant behavior when the individual photon energy
approaches the energy of the discrete exciton state, resulting in
a satisfactory description of the data. When the photon energy
exceeds the band gap, the role of the bound exciton states
decreases and instead the response is largely dictated by
continuum states, which results in a slight decrease of
$\alpha_2$. Note that only the dephasing times in the RDMA
analysis are fitting parameters, while all other parameters are
obtained from the tabulated values shown Tables
\ref{Sidata1}-\ref{Sidata3}.

\begin{figure}[H]
\includegraphics[width=0.5\textwidth]{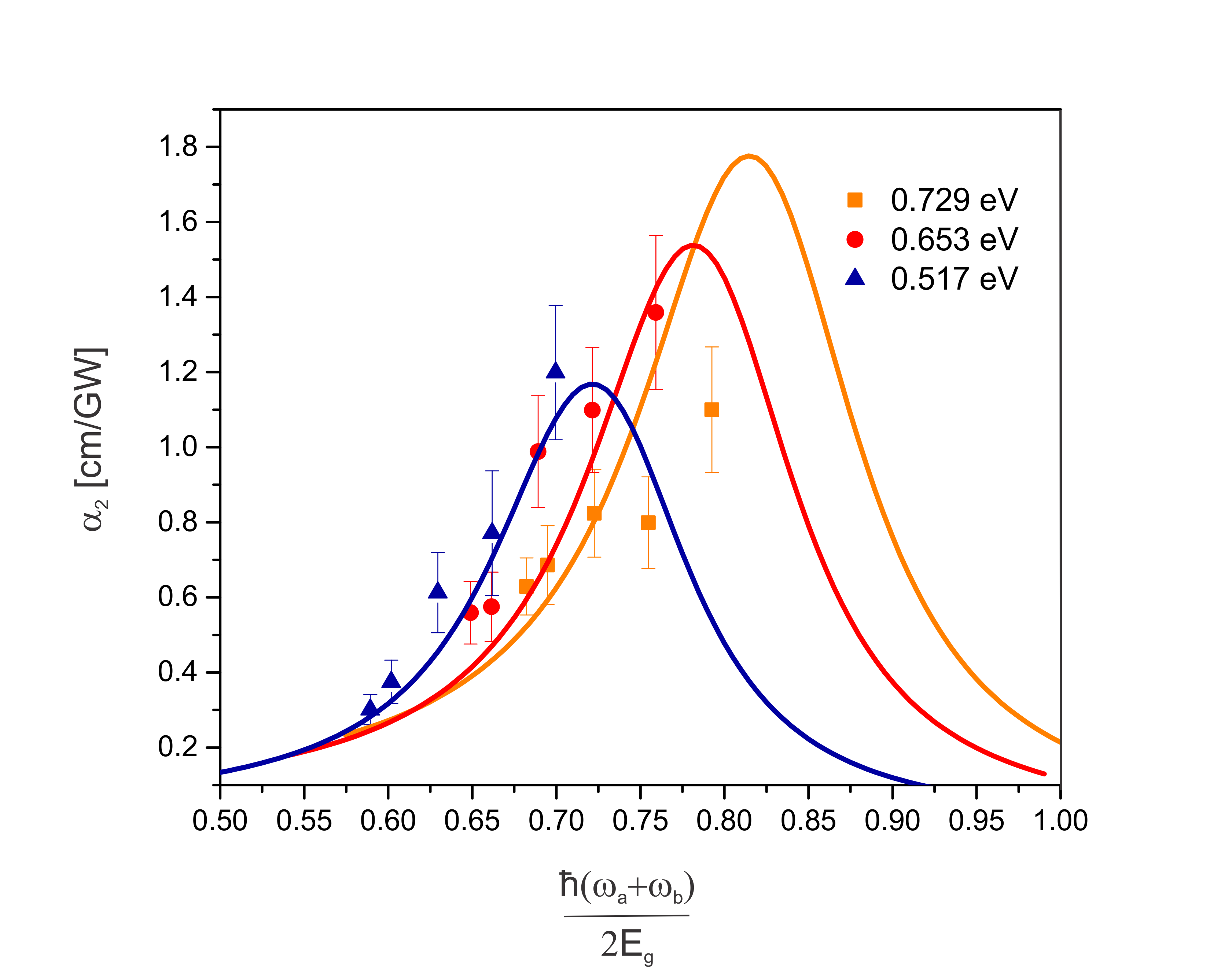}
\caption{\label{fig:creol} Scaling of the non-degenerate two-photon absorption coefficient $\alpha_2$ as a function of the normalized equivalent energy. Experimental data points are obtained from Ref~\cite{faryadras_non-degenerate_2021} and thick solid lines represent the results from the RDMA analysis. Different colors correspond to different photon energies $\hbar\omega_a$.}
\end{figure}

\begin{figure}[H]
\includegraphics[width=0.5\textwidth]{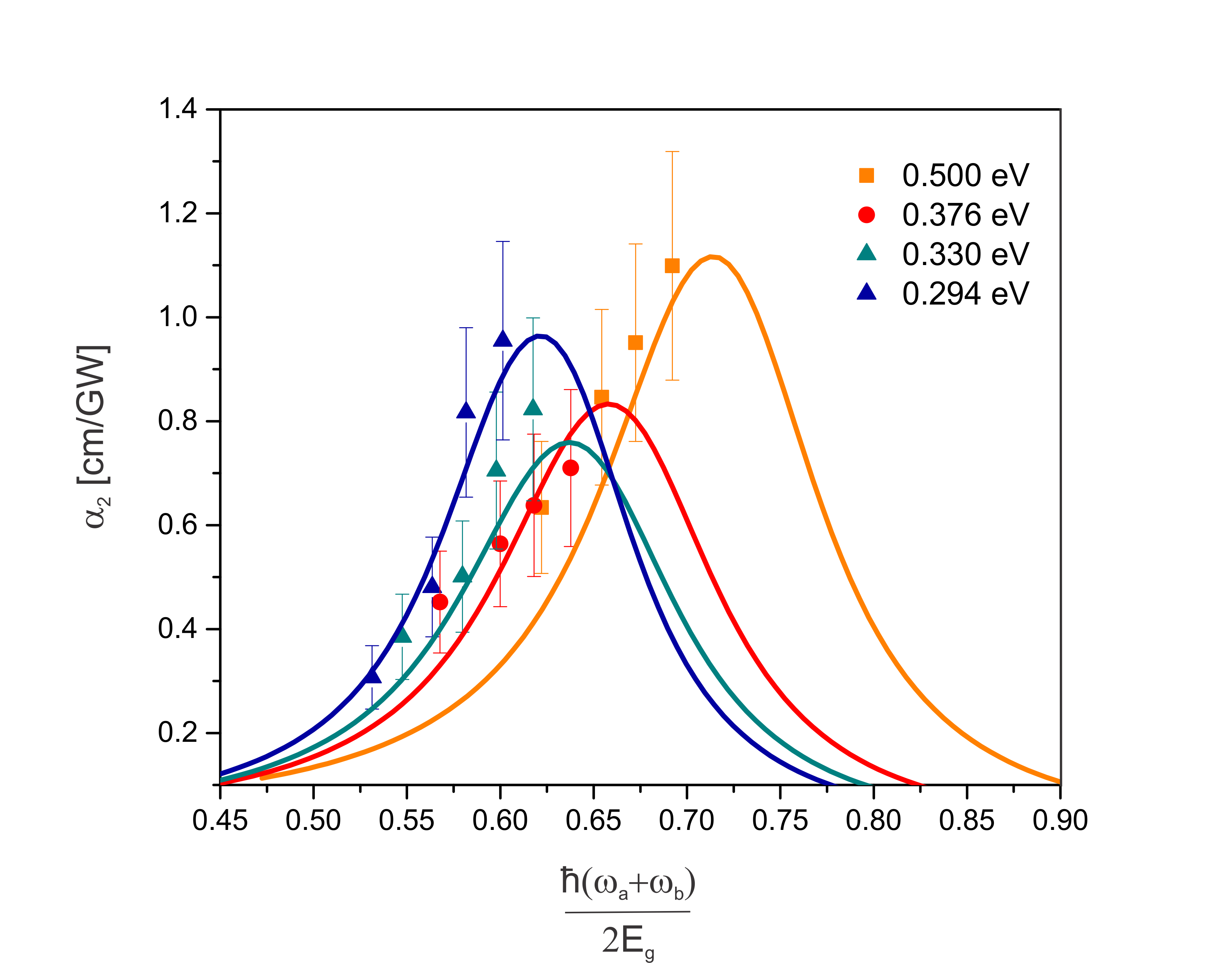}
\caption{\label{fig:UCI} Scaling of the non-degenerate two-photon absorption coefficient $\alpha_2$ as a function of the normalized equivalent energy. Data points indicate experimental results obtained in current study and thick solid lines represent the results from the RDMA analysis. Different colors correspond to different photon energies $\hbar\omega_a$.}
\end{figure}

We next use the model to describe the energy scaling of the NTA
coefficient as measured in
Ref~\cite{faryadras_non-degenerate_2021}, which encompasses photons with energy ratio from 1.6 to 2.08 (Figure \ref{fig:creol}). This compliments our study using pump
photon energies of 0.294~eV--0.5~eV and probe energies of
0.91~eV--1.08~eV, corresponding to photon energy ratio from 1.82 to 3.67 (Figure \ref{fig:UCI}). The theoretical
predictions for $\alpha_2$ are in good agreement with both
experimental data sets. The most notable feature in the normalized
photon energy curve is the appearance of a peak, which the model
attributes to the presence of the bound exciton resonance. The
position of this peak shifts in accordance with the energy tuning
of the lower energy pump photon, and its magnitude is largely
dictated by the damping terms in the resonant denominators of
$\alpha_{2dH}$. However, this resonant behavior is significantly
different from what is reported
in~\cite{faryadras_non-degenerate_2021}, where the theoretical
predictions made anticipate a quasi-linear scaling of $\alpha_2$
as a function of the normalized photon energy. Future work may
include NTA experiments to verify the behavior of
$\alpha_2$ and the influence of continuum states above the band
edge, such as can be obtained with a two-color Z-scan.

This work is a significant departure from the conventional
understanding of the two-photon absorption process in
semiconductors. Previous methods have described the 2PA absorption
process in terms of transition rate matrix elements based on
electron band states, where significantly non-degenerate photons
become resonant with the interband transition and an intraband
transition simultaneously. Within the framework of the RDMA, the
defining feature of the absorption process is the ability to
produce either bound or free excitons within the material.
Tracking the formation of excitons allows the RDMA to yield
general expressions for semiconducting materials with low binding
energies, i.e. only the lowest discrete state needs to be
considered. These results are analytical in that there is no
parametrization done to provide free variables for fitting. The
only variability results from loosely determined material
parameters, namely the relaxation times $T_1$ and $T_2$.

It is possible to apply this analysis to direct gap two-photon
transitions as well, which omits the involvement of phonon
absorption or emission processes in the final equations. The RDMA
also accounts for the physical realities of material anisotropy
and relaxation mechanisms that are often added phenomenologically
elsewhere. Bolstered by the excellent agreement between theory and
experiment over a wide range of photon energies and non-degeneracy
ratios, we believe that the current description provides a deeper
insight into two-photon absorption process in semiconducting
materials.

\section{Conclusion}

We have employed the RDMA approach to elucidate two-photon absorption (2PA) in indirect gap semiconductors, using silicon as the representative material. Our methodology, in contrast to existing models for 2PA in semiconductors, uniquely incorporates a detailed description of DTA (degenerate two-photon absorption) and NTA (non-degenerate two-photon absorption) through the production of excitons, a physical insight often overlooked in prior analyses. The approach allows the computation of the non-degenerate two-photon absorption coefficient, capturing the influence of both bound and free excitonic states within the material. In addition, the RDMA enables the inclusion of the effects of phonon-assisted transitions and material anisotropy. According to our model, the two-photon absorption process in silicon intensifies as the energy of individual incident photons, while below the gap, approaches the single bound excitonic state. This prediction aligns with recently obtained experimental NTA absorption data, shedding new light on the interpretation of published DTA data in bulk silicon. Above the gap, the 2PA process primarily occurs through the continuum of free states, providing a satisfactory description of the DTA data across a wide range of photon energies  ($\hbar\omega/E_g=0.5-1.3$). Last, but not least, the analysis outlined using the RDMA can be generalized and readily applied to other indirect gap materials by adjusting the necessary material values. Furthermore, the model can be adapted to describe direct transitions by excluding the involvement of phonon modes to complete the transition.

\begin{acknowledgments}
DAF and EOP acknowledge support of the Chan-Zuckerberg Initiative and the National Health Institute, grant R21-GM141774. AS acknowledges the support of the Air Force Office of Scientific Research (AFOSR) through the grant FA9550-23-1-0267.
\end{acknowledgments}

\section*{Data Availability Statement}
The data that support the findings of this study are available from the corresponding author upon reasonable request.

\section*{Bibliography}

\bibliography{sirdma2}

\end{document}